\documentclass{article}
\usepackage{arxiv}

\usepackage[utf8]{inputenc} 
\usepackage[T1]{fontenc}    
\usepackage{hyperref}       
\usepackage{url}            
\usepackage{booktabs}       
\usepackage{amsfonts}       
\usepackage{nicefrac}       
\usepackage{microtype}      
\usepackage{lipsum}		
\usepackage{graphicx}
\usepackage{cite}

\usepackage{array}
\usepackage{amssymb, amsmath, amsthm}
\usepackage{graphicx}
\usepackage{lmodern,url}
\usepackage{makecell} 
\usepackage{cancel}
\usepackage{multirow}
\usepackage{microtype}
\usepackage{lineno}
\usepackage{xspace}
\usepackage{xcolor}
\usepackage{siunitx}
\usepackage{todonotes}
\usepackage{booktabs}
\usepackage[ruled,vlined]{algorithm2e}
\usepackage{optidef}
\usepackage{mathdots}
\usepackage{subcaption}
\usepackage{csquotes}
\usepackage{caption}
\newcommand{\dis}{\displaystyle}

\DeclareMathOperator{\sgn}{sgn}

\newcommand{\figref}[1]{Fig.~\ref{#1}}
\newcommand{\tabref}[1]{\tablename~\ref{#1}}
\graphicspath{{Figures/}}

\usepackage{amsmath}
\usepackage{tikz}
\usetikzlibrary{fit}
\usetikzlibrary{positioning}
\tikzset{%
  highlight/.style={rectangle,rounded corners,fill=red!15,draw,fill opacity=0.25,thick,inner sep=0pt}
}

\definecolor{mygreen}{RGB}{160, 242,182}

\newcommand{\xunderbrace}[2][\vphantom{\dfrac{A}{A}}]{\underbrace{#1#2}}
\definecolor{mypurple}{RGB}{236, 223, 234}

\definecolor{knBlue}{RGB}{210,230,255}

\newcommand{\Ieff}{\ensuremath{I^{\rm eff}}\xspace}
\newcommand{\INF}{\ensuremath{\mathcal{I}}\xspace}
\newcommand{\CM}{\ensuremath{C_{ji}}}
\newcommand{\CMij}{\ensuremath{C_{ij}}}
\newcommand{\kseason}{\ensuremath{k_{\rm seasonality}}\xspace}

\newcommand{\IB}{\ensuremath{I^B}\xspace}
\newcommand{\EBn}{\ensuremath{E^{n}}\xspace}
\newcommand{\IBn}{\ensuremath{I^{n}}\xspace}
\newcommand{\EBv}{\ensuremath{E^{v}}\xspace}
\newcommand{\IBv}{\ensuremath{I^{v}}\xspace}
\newcommand{\ICU}{\ensuremath{{\rm ICU}}\xspace}
\newcommand{\avICU}{\ensuremath{H}\xspace}
\newcommand{\latRate}{\ensuremath{\rho}\xspace}
\newcommand{\fracImmun}{\ensuremath{\eta}\xspace}

\newcommand{\avProtection}{\ensuremath{\kappa}\xspace}

\allowdisplaybreaks

\title{Interplay between risk perception, behaviour, and COVID-19 spread}

\usepackage{authblk}

\author[1$\dagger$]{Philipp Dönges}
\author[1$\dagger$]{Joel Wagner}
\author[1,2$\dagger$]{Sebastian Contreras}
\author[1$\dagger$]{Emil Iftekhar}
\author[1]{Simon Bauer}
\author[1]{Sebastian B. Mohr}
\author[1]{Jonas Dehning}
\author[3]{Andr\'e Calero Valdez}
\author[4]{Mirjam Kretzschmar}
\author[5]{Michael M\"as}
\author[6]{Kai Nagel}
\author[1,7*]{Viola Priesemann}

\affil[1]{Max Planck Institute for Dynamics and Self-Organization, G\"ottingen, Germany.}
\affil[2]{Centre for Biotechnology and Bioengineering, Universidad de Chile, Beauchef 851, 8370456 Santiago, Chile.}
\affil[3]{RWTH Aachen University, Aachen, Germany.}
\affil[4]{University Medical Center Utrecht, Utrecht, The Netherlands.}
\affil[5]{Karlsruhe Institute of Technology, Karlsruhe, Germany.}
\affil[6]{Technische Universität Berlin, Berlin, Germany.}
\affil[7]{Institute for the Dynamics of Complex Systems, University of G\"ottingen,  G\"ottingen, Germany.}
\affil[ ]{{$*$} Corresponding Author: Viola Priesemann (viola.priesemann@ds.mpg.de)}
\affil[ ]{{$\dagger$} These authors contributed equally}

\date{}

\renewcommand{\headeright}{ }
\renewcommand{\undertitle}{ }

\begin{document}
\maketitle

\begin{abstract} 
Pharmaceutical and non-pharmaceutical interventions (NPIs) have been crucial for controlling COVID-19. 
They are complemented by voluntary health-protective behaviour, building a complex interplay between risk perception, behaviour, and disease spread. 
We studied how voluntary health-protective behaviour and vaccination willingness impact the long-term dynamics.
We analysed how different levels of mandatory NPIs determine how individuals use their leeway for voluntary actions. 
If mandatory NPIs are too weak, COVID-19 incidence will surge, implying high morbidity and mortality before individuals react;
if they are too strong, one expects a rebound wave once restrictions are lifted, challenging the transition to endemicity.
Conversely, moderate mandatory NPIs give individuals time and room to adapt their level of caution, mitigating disease spread effectively. When complemented with high vaccination rates, this also offers a robust way to limit the impacts of the Omicron variant of concern. 
Altogether, our work highlights the importance of appropriate mandatory NPIs to maximise the impact of individual voluntary actions in pandemic control.
\end{abstract}

\clearpage
\section*{Introduction}
During the COVID-19 pandemic, the virus has played a central role in people's day-to-day conversations and the information they search for and consume \cite{Casero2020info}. The growing amount of news and specialised literature on COVID-19 can inform individual decisions in a wide range of situations and on various timescales \cite{kim2020info}. For example, people decide multiple times every day how closely they follow mask-wearing regulations or meeting restrictions. However, if hesitant, they might take weeks or months to decide whether to accept a vaccine. These decisions impact the spreading dynamics of COVID-19 and ultimately determine the effectiveness of interventions and how smoothly we transit to SARS-CoV-2 endemicity.

While typical models of disease spread consider that individual behaviour affects the spreading dynamics of an infectious disease, they often neglect that there is also a relation in the opposite causal direction. This feedback loop comprises that, e.g., mass media regularly updates individuals on the latest local developments of the pandemic, such as the current occupancy of intensive care units (ICUs). This information affects individuals' opinions and risk perceptions and, thus ultimately their actions \cite{ferrer2015risk}. For example, given high perceived risk, individuals reduce their non-essential contacts beyond existing regulations and increase their willingness to accept vaccine offers accordingly, an effect observed in empirical research conducted with routine surveys in Germany \cite{betsch2020monitoring} and other parts of the world \cite{imbriano2021online,perrotta2021behaviours,druckman2021affective,salali2021effective}. However, to quantify the effect of individual voluntary actions on the dynamics of COVID-19, two questions remain open: 
(1) What is the relationship between risk perception and voluntary action, on the one hand, and the spread of the disease, on the other hand; and 
(2) what is the relative contribution of voluntary action when mandatory restrictions are in place? 

\begin{figure}[!ht]
    \centering
    \includegraphics[width=120mm]{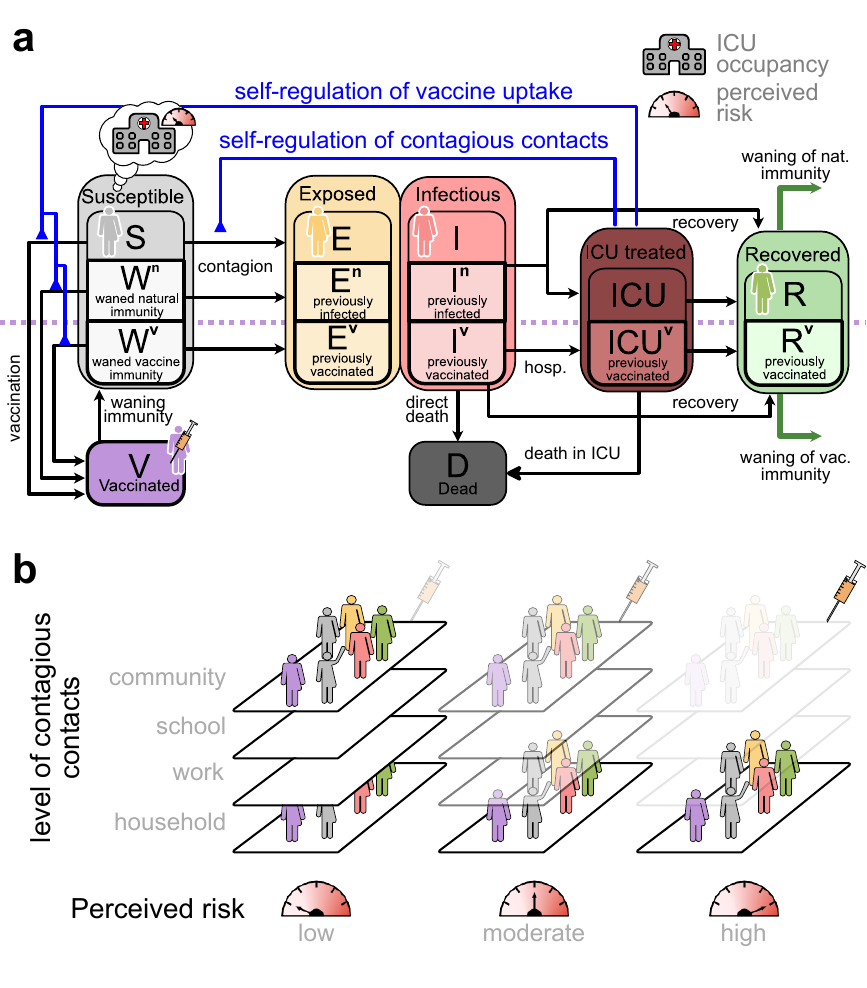}
    \caption{%
        \textbf{Interplay between risk perception and voluntary health-protective behaviour.} \textbf{a:} Sketch of the proposed age-stratified compartmental model of disease spread, which incorporates different stages for disease progression and immunological conditions of the susceptible population with their respective chances of being infected and developing a severe course (Fig.~S1, Supplementary Information, for full model). The behavioural feedback (blue lines) changes individuals' contagious contact behaviour, as well as their willingness to get vaccinated, and hence the effective spreading rate. 
        \textbf{b:} We use the contact matrix of \cite{mistry2021inferring}, which yields the contact rates at home, school, work and in the community for each age-group. For the subsequent scenarios, we adapt these contexts of contacts separately. Some of the contacts are by definition hard to reduce voluntarily (e.g., household contacts), while others (at school and work) strongly depend on current mandatory non-pharmaceutical interventions (Fig.~S3 for details).
        }
    \label{fig:Figure_1}
\end{figure}

In this work, we aim to quantify the impact of voluntary actions on disease spread while studying the questions mentioned above for the COVID-19 pandemic. 
(1) We analyse survey and COVID-19 vaccination data in European countries to uncover the relationship between the occupancy of ICUs---which determines the perceived risk---and voluntary immediate health-protective behaviour as well as the willingness to get vaccinated. We then incorporate these effective feedback loops into a deterministic compartmental model (\figref{fig:Figure_1}a). 
(2) We decompose the overall contact structure into contextual contacts (\figref{fig:Figure_1}b) and for each context define a range in which voluntary action can be adapted according to individual risk-perception, given the level of mandatory non-pharmaceutical interventions (NPIs). To that end, we use the functional form identified in (1) (\figref{fig:Figure_2}). 
We explore different intervention scenarios in the face of adverse seasonality \cite{Gavenciak2021seasonality,moriyama2020seasonality,sajadi2020temperature}, using as reference the winter 2021/2022 in central Europe. Our analysis confirms that both extremes ('freedom day' or stringent measures throughout) bear large harms in the long run. However, when measures leave space for voluntary actions, people's adaptive behaviour can efficiently contribute to breaking the wave and change the course of the pandemic. 


\section*{Results}

\subsection*{Data-derived behavioural feedback loops}

Throughout this manuscript, we investigate how the interplay between information about the COVID-19 pandemic and its spreading dynamics is mediated by the perception of risk. Risk perception modulates both, (i) people's immediate voluntary health-protective behaviour, e.g., their level of contacts and their adherence to mask-wearing and hygiene recommendations, and (ii) their willingness (or hesitancy) to receive vaccination (Fig.~\ref{fig:Figure_1}). Individuals constantly receive information on the current COVID-19 incidence, ICU occupancy, and deaths (which are all closely related \cite{olivieri2021covid,bravata2021association,Linden2020DAE}) either via news outlets or because of reports about COVID-19 cases in their social circles. Hence, the risk they perceive depends on this evolving trend over time. 

We tailor our approach to the situation of the COVID-19 pandemic, i.e., to a disease having the following characteristics: (i) high transmissibility, (ii) relatively low infection fatality rate, (iii) widespread vaccine hesitancy, (iv) waning immunity, and (v) public attention and coverage. We differentiate from the approaches of \cite{epstein2008coupled, epstein2021triple, bauch2005imitation} as we neither model the contagion of fear explicitly nor a direct coupling between incidence and fear. Instead, we assume that individuals build their perception of risk based on the ICU occupancy over time using a memory function, similar to the theoretical approach in \cite{d2009information, d2007vaccinating}. This is a sensible choice, as ICU occupancy signals i) how likely governmental bodies are to re-implement emergency NPIs to prevent overwhelming healthcare facilities (and thereby limit individual freedoms), and ii) how likely it is that an individual's close contacts (or their contacts) would have been severely ill. Besides, our modelling framework constitutes a methodological advancement from that presented in \cite{epstein2021triple}, as we provide a detailed description of all epidemiologically relevant disease states and several external effects influencing its spread, such as seasonality, contextual contact networks and NPIs.

We assume that individuals base their decisions about heath-protective behaviour on the recent developments of the pandemic. Following the ideas of Zauberman et al. about perception of time in decision-making \cite{zauberman2009discounting}, we consider that when individuals decide about behaviour that only has immediate protective effects, they consider only the current risk-level. For instance, when deciding whether or not to wear a mask in the supermarket on a given day, they only consider the most recently reported ICU occupancy. Decisions with longer-term protection, in contrast, are also based on a longer-term risk-assessment. When deciding whether or not to get a booster vaccine, for example, individuals do not only take into account the ICU-occupancy on the day of the decision but they are looking back at a longer period. We detail the assumptions about the perceived risk-level and the resulting health-protective behaviour in the Methods section. In the following, we sketch the derivation of the feedback loops from this perceived risk to people's immediate voluntary health-protective behaviour and willingness to get vaccinated.

\begin{figure}[!ht]
    \centering
    \includegraphics[width=6.5in]{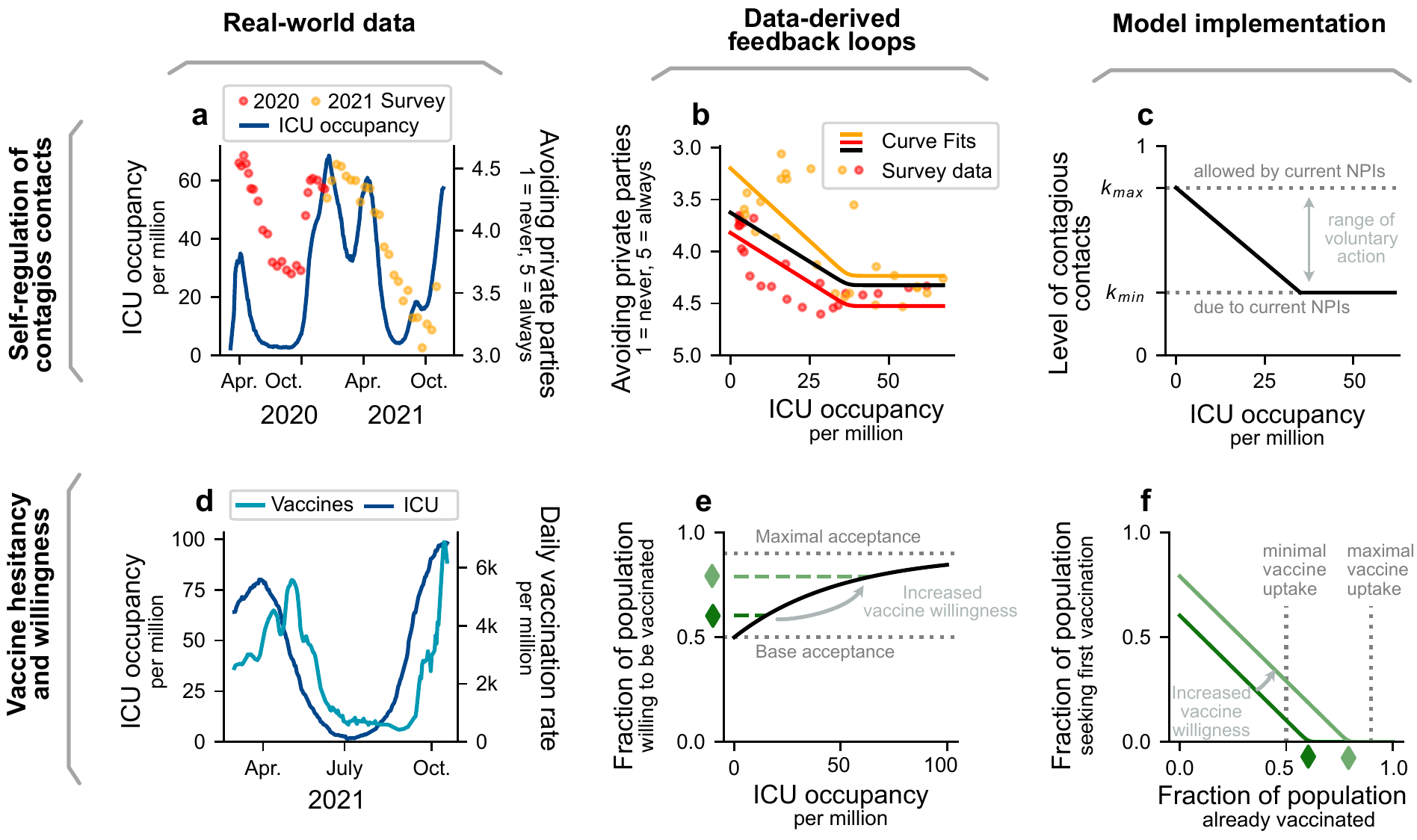}
    \caption{%
        \textbf{Data-derived formulation of behavioural feedback loops.} \textbf{a}: Reported contact reductions follow intensive care unit (ICU) occupancy in Germany. Survey participants were asked how likely they were to avoid private parties over the course of the pandemic on a discrete scale from 1 (never) to 5 (always) \cite{betsch2020monitoring}. To decouple the effect of vaccination availability, we present 2020 (red) and 2021 (yellow) data separately. Ticks indicate the middle of the month. \textbf{b}: The survey data on contact reduction and the ICU occupancy are related. The piece-wise linear relationship shows the reduction of contacts with increasing ICU occupancy, and for even higher ICU occupancy a saturation. Red, yellow, and black  represent fits to the data from 2020, 2021, and overall, respectively. \textbf{c}: In the model, the contact reduction and its dependency on ICU occupancy is implemented as a multiplicative reduction factor $k$ that weighs the age-dependent contextual contact matrices (\figref{fig:Figure_1}b). \textbf{d}: Vaccine uptake increases with ICU occupancy in Romania (shown here) and other European countries (Fig.~S4). \textbf{e}: Willingness to accept a vaccine offer is modelled using an exponentially-saturating function, ranging between a lower and upper bound of acceptance depending on ICU occupancy. The bounds represent that a fraction of people is willing to be vaccinated even at no immediate threat (no ICU occupancy), and another fraction is not willing or able to get vaccinated no matter the threat. \textbf{f}: Vaccines are delivered at a rate proportional to the number of people seeking a vaccine, i.e., the difference between the number of people willing to be vaccinated and those already vaccinated. Thus, when the number of already vaccinated equals the number of people willing to get vaccinated, no more vaccinations are carried out. The same functional shape describes the booster uptake.
        }
    \label{fig:Figure_2}
\end{figure}

\textbf{Feedback on health-protective behaviour}

To determine the explicit relationship between the perceived level of risk and immediate voluntary health-protective behaviour---which presents one of the feedback loops in our model---we exploit results from the German COSMO study, a periodic survey where participants are asked about their opinions and behaviour regarding the COVID-19 pandemic and NPIs \cite{betsch2020monitoring}. Their answers on adhering to health-protective behaviour recommendations (avoiding private parties in this case) correlate with the ICU occupancy in Germany at the time (Fig.~\ref{fig:Figure_2}a). However, at very high ICU occupancy, adoption of health-protective behaviour seems to reach a plateau (Fig.~\ref{fig:Figure_2}b); no further adoption seems to be feasible, arguably because those individuals willing to engage in health-protective behaviour have done so already as far as they can, and those unwilling are insensitive to higher burden on ICUs.
Hence, we fit a piece-wise linear function (with a rounded edge at the transition - called a softplus) to the COSMO data (Pearson correlation coefficient r=0.64 for 2020-2021 (black), r=0.81 for 2020 (red) and r=0.53 for 2021 (yellow)) and use it for the feedback between information in terms of ICU occupancy and voluntary health-protective behaviour (Fig.~\ref{fig:Figure_2}c and Methods for details).

\textbf{Feedback on vaccination behaviour}

The second feedback loop in our model describes the relationship between the level of perceived risk and vaccine hesitancy. To quantify it, we study the vaccination trends in different European countries and compare them with the trends in ICU occupancy (Fig.~S4, Supplementary Information). The case of Romania (Fig.~\ref{fig:Figure_2}d) illustrates the relation very clearly: Vaccination rates follow the ICU occupancy with a delay of a few weeks. By analysing the correlation between vaccination rate and ICU occupancy with a variable delay, we reach the highest Pearson correlation coefficient (0.96) with a delay of 25 days. However, the specific reaction delay and magnitude of the effect differs between countries (Fig.~S4). In our model, we propose that as ICU occupancy increases so does the willingness to get vaccinated (i.e., higher probability of accepting a vaccine offer when ICU occupancy is high). As not everybody in the population is willing to accept a vaccine offer, the willing fraction of the population is a function that saturates below 1 (Fig.~\ref{fig:Figure_2}e). With this formulation, vaccinations are only carried out if the fraction of the population willing to get vaccinated is larger than the fraction of currently vaccinated (Fig.~\ref{fig:Figure_2}f and Methods for details). 

Our model can capture two features observed in real-world vaccination programmes. First, when case numbers are low and vaccine uptake high, rational agents might have insufficient incentives for getting vaccinated. Assuming a high perceived risk of vaccine side effects, the agents would thus decline vaccination when offered. The above is known as the free-rider problem in game theory and economics \cite{bauch2012evolutionary}. Second, the two feedback loops in our model and the incorporation of waning immunity allows us to observe different incidence curve shapes and replicate recurrent waves of infections. The above is a necessary validity check, as real-world outbreaks exhibit a large variety of incidence curve shapes \cite{tkachenko2021stochastic}. These may ultimately unveil universal patterns of disease spread that are consistent across countries \cite{dankulov2021worldwide}.

\subsection*{Behavioural feedback loops yield more realistic results than classical models}

Classical SEIR-like compartmental models have found wide application in the first stages of the COVID-19 pandemic. In these models, the different stages of disease progression are represented by separate compartments and individuals transit from one to another at a given (and typically constant) transition rate. In that way, an infectious disease outbreak will proliferate if the spreading rate of the disease is larger than the recovery rate and if a large-enough fraction of the population is susceptible to being infected. However, these simple models often tend to overestimate the size of an infectious disease outbreak or all possible trajectories for the incidence trends \cite{tkachenko2021stochastic}, as they do not incorporate mechanisms of dynamical adaptation of restrictions \cite{bauer2021relaxing} or, as studied in this paper, behaviour.

\begin{figure}[!ht]
\hspace*{-1cm}
    \centering
    \includegraphics[width=85mm]{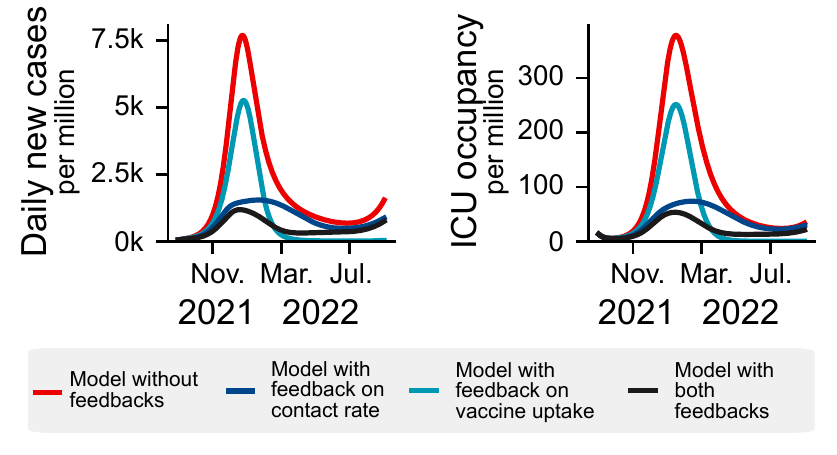}
    \caption{%
        \textbf{Incorporating behavioural feedback loops in compartmental models broadens the dynamic range of the solutions and yields more realistic results.} Different variations of a compartmental model are displayed to show the effect of the two feedback loops used in our model: When ICU occupancy increases, individuals increase their health-protective behaviour and are more willing to be vaccinated. This dynamical adaptation can break a wave at lower case numbers and lead to extended infection plateaus (blue curves), which a classic compartment model is unable to reproduce as it does not incorporate the population's reaction to the disease (red curve). 
        }
    \label{fig:Figure_3}
\end{figure}

We observe that including the feedback loops described above reduces the peak in incidences and hospitalisations while keeping the timing of the wave almost unchanged (see Fig.~\ref{fig:Figure_3}). More generally, these feedback loops break increasing and declining trends, resulting in long but flat infection plateaus or multiple waves. Compared to classical SEIR-like models, where two dynamical regimes are possible ---exponential growth or decay of case numbers, when neglecting waning immunity---, our model captures a broader spectrum of dynamics by linking ICU occupancy with individuals' health-protective voluntary behaviour and vaccine uptake. 

\subsection*{Policies with either too weak or too strong interventions throughout winter bear higher levels of mortality and morbidity}

\begin{figure}[!ht]
\hspace*{-1cm}
    \centering
    \includegraphics[width=7in]{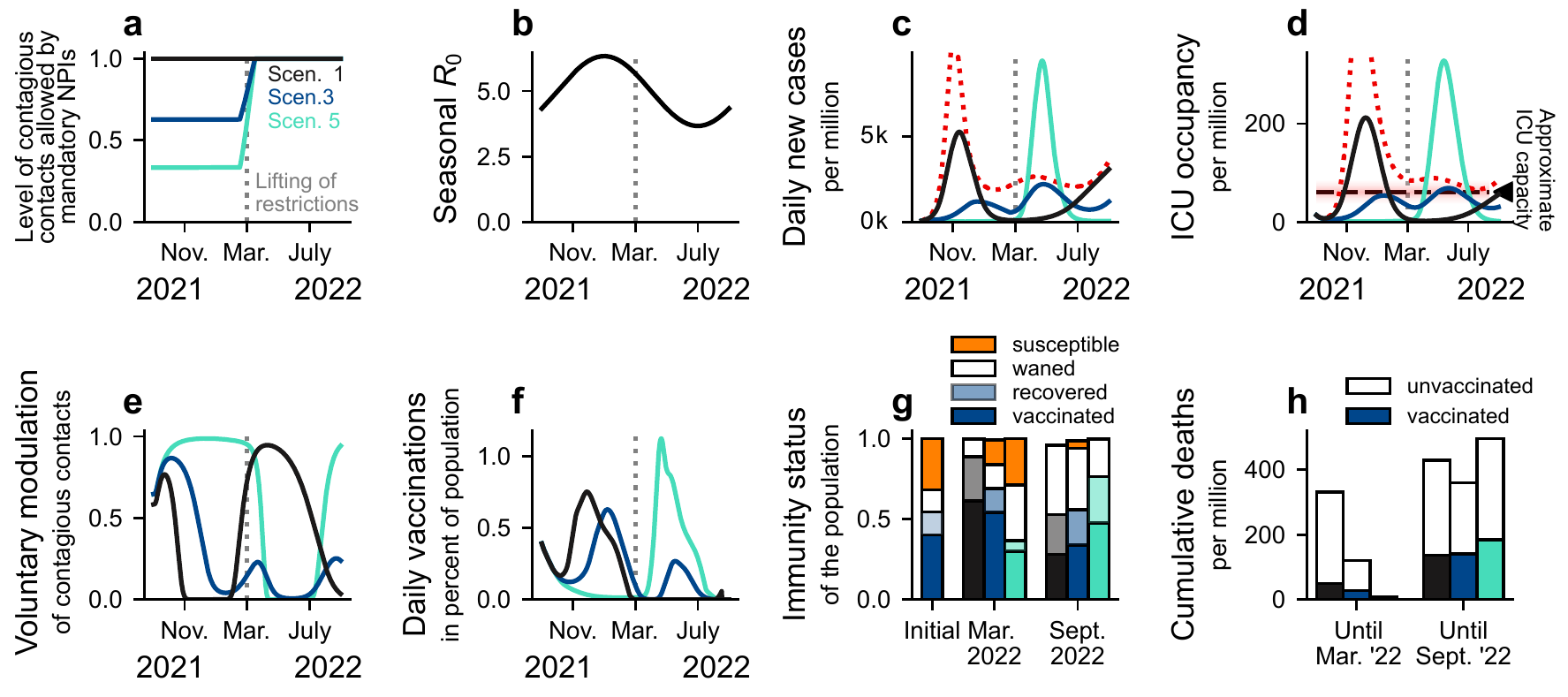}
    \caption{%
        \textbf{Maintaining moderate contact restrictions throughout winter outperforms extreme scenarios in balancing the burden on ICUs by allowing people the freedom to act according to their risk perception.}
        The level of mandatory NPIs sustained throughout winter 2021/2022, together with people's voluntary preventive actions, determines case numbers and ICU occupancy over winter and beyond. Ticks are set on the first day of the month.
        \textbf{a:} The three displayed scenarios of mandatory NPI stringency in winter reflect "freedom-day" with only basic hygiene measures (black), considerable contact reduction and protective measures (e.g., mandatory masks) in school, at the workplace and in the community (blue), and strong contact reduction and partial school closure (mint). All measures are gradually lifted centred around March 1st 2022, over the course of four weeks. 
        \textbf{b:} The seasonality of the basic reproduction number $R_0$. 
        \textbf{c, d:} Scenario 1 (black): Without mandatory restrictions, incidence and ICU occupancy increase steeply; this increases voluntary health-protective behaviour and vaccine uptake in the population (\textbf{e, f}), and leads to higher rates of naturally acquired immunity (\textbf{g}), but also high mortality and morbidity in winter (\textbf{h}). Note that disproportionally more vaccinated individuals die after March 2022 because, at this point, most of the population is vaccinated. A 'full wave' is added in \textbf{c,d} (red dotted line), depicting the development of case numbers and ICU occupancy in the absence of behavioural feedback mechanisms. 
        Scenario 3 (blue): Maintaining moderate restrictions would prevent overwhelming ICUs while allowing for higher vaccine uptakes and rates of post-infection immunity. 
        Scenario 5 (mint): Maintaining strong restrictions would minimise COVID-19 cases and hospitalisations in winter, generating a perception of safety across the population. However, this perceived safety is expected to lower the incentives to get vaccinated. Furthermore, immunity of all kinds will wane over winter. Altogether, this can cause a severe rebound wave if restrictions are completely lifted in March.
        Furthermore, in all scenarios where ICU capacity is exceeded, we would in reality expect either disproportionally higher mortality due to the burden on the health system or a change in mandatory NPIs.
        }
    \label{fig:Figure_4}
\end{figure}

Using parameters obtained from surveys and other data sources (Table~S3, Supplementary Information), we analyse five scenarios of mandatory NPIs throughout winter (for all age-stratified results see Supplementary Information): 1) no NPIs at all, 2)-4) moderate NPIs and 5) strong NPIs (Methods for details). The stringency of the scenarios and the seasonal effects are depicted in \figref{fig:Figure_4}a, b and \figref{fig:Figure_5}a, b. As an example case, we assume a country with a total vaccination rate of $60\%$ and a recovered fraction of $20\%$. Note that we include the possibility of overlaps between vaccinated and recovered. Thus, the total fraction of immune individuals does not add up to $80\%$ but $68\%$. For more detail on the initial conditions, see Supplementary Information, Sec.~S3.1. 

Without any mandatory NPIs throughout winter (Scenario 1, Fig.~\ref{fig:Figure_4}, black lines), case numbers and hospitalisations will show a steep rise (Fig.~\ref{fig:Figure_4}c,d). As a consequence, individuals voluntarily adapt their health-protective behaviour and are more inclined to accept a vaccine offer (Fig.~\ref{fig:Figure_4}e--g). Although this scenario features unrealistically high mortality and morbidity, modelling results in the absence of any behaviour feedback mechanisms yield even higher levels (cf. Fig~\ref{fig:Figure_4}c, d, dotted red line). 

In contrast, suppressing the seasonal wave through strong mandatory NPIs (Scenario 5, Fig.~\ref{fig:Figure_4}, mint lines) and thereby maintaining low case numbers through winter only delays the wave to a later but inevitable date once restrictions are lifted (Fig.~\ref{fig:Figure_4}c,d). Low COVID-19 incidence throughout winter implies i)~low post-infection immunity, ii)~little incentives for first or booster vaccination, iii) waning immunity, and iv)~lower rates of "naturally" boosting immune memory upon re-exposure to the virus \cite{brown2021original}. The resulting low immunity levels (Fig.~\ref{fig:Figure_4}g) then fuel a higher rebound wave when restrictions are lifted in March 2022, despite favourable seasonality. Similar rebound waves have been observed for other seasonal respiratory viruses \cite{gomez2021uncertain,sanz2021social}.

Interestingly, the middle strategy, namely moderate NPIs during winter, prevents the high wave in winter as well as the rebound wave in spring that characterise the scenarios with no or with strong NPIs, respectively (Scenario 3, Fig.~\ref{fig:Figure_4}, dark blue). Unlike in the extreme scenarios, the ICU capacity in Scenario 3 is not exceeded in any season, hence avoiding reduced health care quality and strong burden to health care workers. Fig.~\ref{fig:Figure_4}h shows that the death toll in Scenario 3 is lower than in the other scenarios. In reality however, this difference would be much larger because scenarios 1 and 5 surpass the assumed ICU capacity by far; that would imply disproportionally higher mortality, an effect we did not quantify in our model. Alternatively, emergency mandatory NPIs would be introduced, which we do not model here.

\subsection*{Voluntary actions can dampen the wave if restrictions are moderate}

\begin{figure}[!ht]
\hspace*{-1cm}
    \centering
    \includegraphics[width=7in]{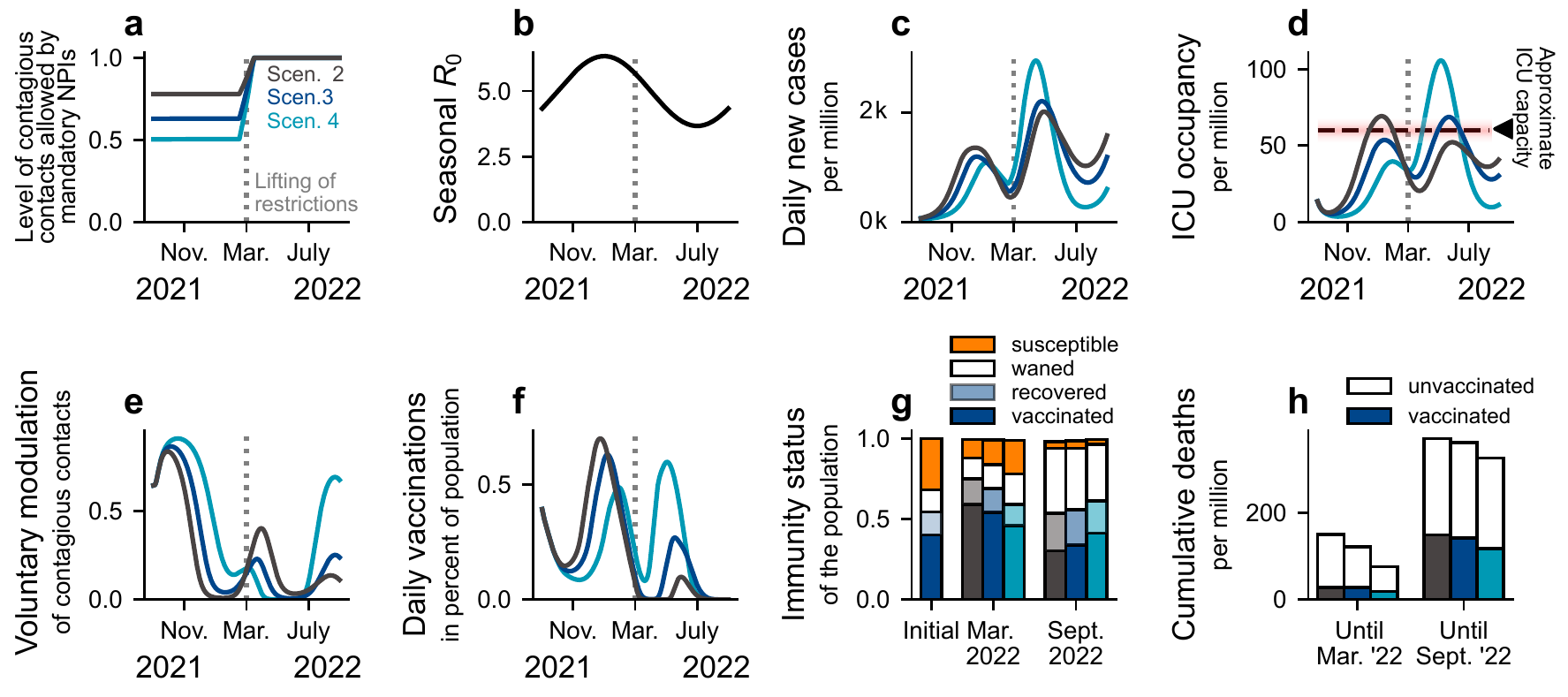}
    \caption{%
        \textbf{Moderate restrictions leave enough room for effective adaptation of behaviour to perceived risk.} \textbf{a:} We explore three scenarios with similar levels of moderate mandatory NPIs sustained throughout winter, the period of adverse seasonality (\textbf{b}). Considering Scenario 3 as reference, moderate restrictions seem to be robust against relaxations of NPIs, as both morbidity and mortality are similar to that of Scenario 2 (\textbf{c}, \textbf{d}, \textbf{h}). However, a perturbation with less strength in the opposite direction (Scenario 4, increasing mandatory NPIs) has a disproportional effect on ICU occupancy. These differences are based on the modulation of voluntary contacts (\textbf{e}) and vaccine uptake (\textbf{f}). Thus, when leaving room for adaption of health-protective measures to perceived risk, people's behaviour will stabilise moderate scenarios where mandatory NPIs are strong enough to prevent a major surge, but not over-protective, so individuals find it rewarding to be vaccinated and to adapt their level of contacts. Note that disproportionally more vaccinated individuals die after March 2022 because, at this point, most of the population is vaccinated (\textbf{g}).
        }
    \label{fig:Figure_5}
\end{figure}

As presented in the previous section, extreme scenarios (Scenarios 1 and 5) bear high levels of morbidity and mortality. However, in scenarios with intermediate restriction levels (Scenarios 2--4, Fig.~\ref{fig:Figure_5}a), voluntary preventive actions (Fig.~\ref{fig:Figure_5}e) can compensate for slightly too low levels of mandatory NPIs, provided that these NPIs are strong enough to prevent a surge in COVID-19 incidence that might be too sudden or strong for individuals to voluntarily adopt health-protective behaviour (Fig.~\ref{fig:Figure_5}c, d). For example, while having different levels of mandatory NPIs, Scenarios 2 and 3 reach similar peaks in ICU occupancy (Fig.~\ref{fig:Figure_5}d). Conversely, despite considering a proportional increase in the strength of NPIs (comparable to that from scenario 2 to 3, Fig.~\ref{fig:Figure_5}a), Scenario 4 is too protective: there are too few incentives to get vaccinated (Fig.~\ref{fig:Figure_5}f) due to the low risk perception as well as too few infections (Fig.~\ref{fig:Figure_5}c) and, hence, appropriate immunity levels are not reached (Fig.~\ref{fig:Figure_5}g). As a consequence, a disproportionally larger off-seasonal wave in spring overwhelms ICUs (Fig.~\ref{fig:Figure_5}d). Noteworthy, even though the nominal mortality is the lowest for Scenario 4 (Fig.~\ref{fig:Figure_5}h), this value does not account for triage-induced over-mortality or novel necessary NPIs that would be likely be imposed and is thus invalid.

\subsection*{Case study: emergence of the Omicron variant of concern and its effect on case numbers}

A risk that cannot be neglected is the emergence of SARS-CoV-2 variants of concern (VOC), such as the Omicron VOC. This variant is rapidly replacing the Delta VOC, thus posing an imminent risk. Although there is substantial uncertainty about its epidemiological features, preliminary evidence shows: Compared to the Delta VOC, Omicron exhibits (i) an increased risk of reinfection or break-through infection \cite{viana2021rapid,pulliam2021omicron_increased,ferguson2021growth}, (ii) a substantial reduction in antibody neutralisation \cite{cele2021omicron_escape,wilhelm2021omicron_reduced,cameroni2021broadly,roessler2021sars,hoffmann2021omicron, gardner2021estimates,perez2021dominican}, (iii) a reduction in vaccine effectiveness against infection \cite{gruell2021mrna,kuhlmann2021breakthrough,nemet2021third,basile2021improved,gardner2021estimates,ferguson2021growth,garcia2021mrna,andrews2021effectiveness}, and (iv) faster spread \cite{torjesenn2021omicron_doublingtime,pulliam2021omicron_increased,ferguson2021growth,barnardmodelling} mainly due to immune escape \cite{lyngse2021sars}.

 \begin{figure}[!ht]
 \hspace*{-1cm}
     \centering
     \includegraphics[width=7in]{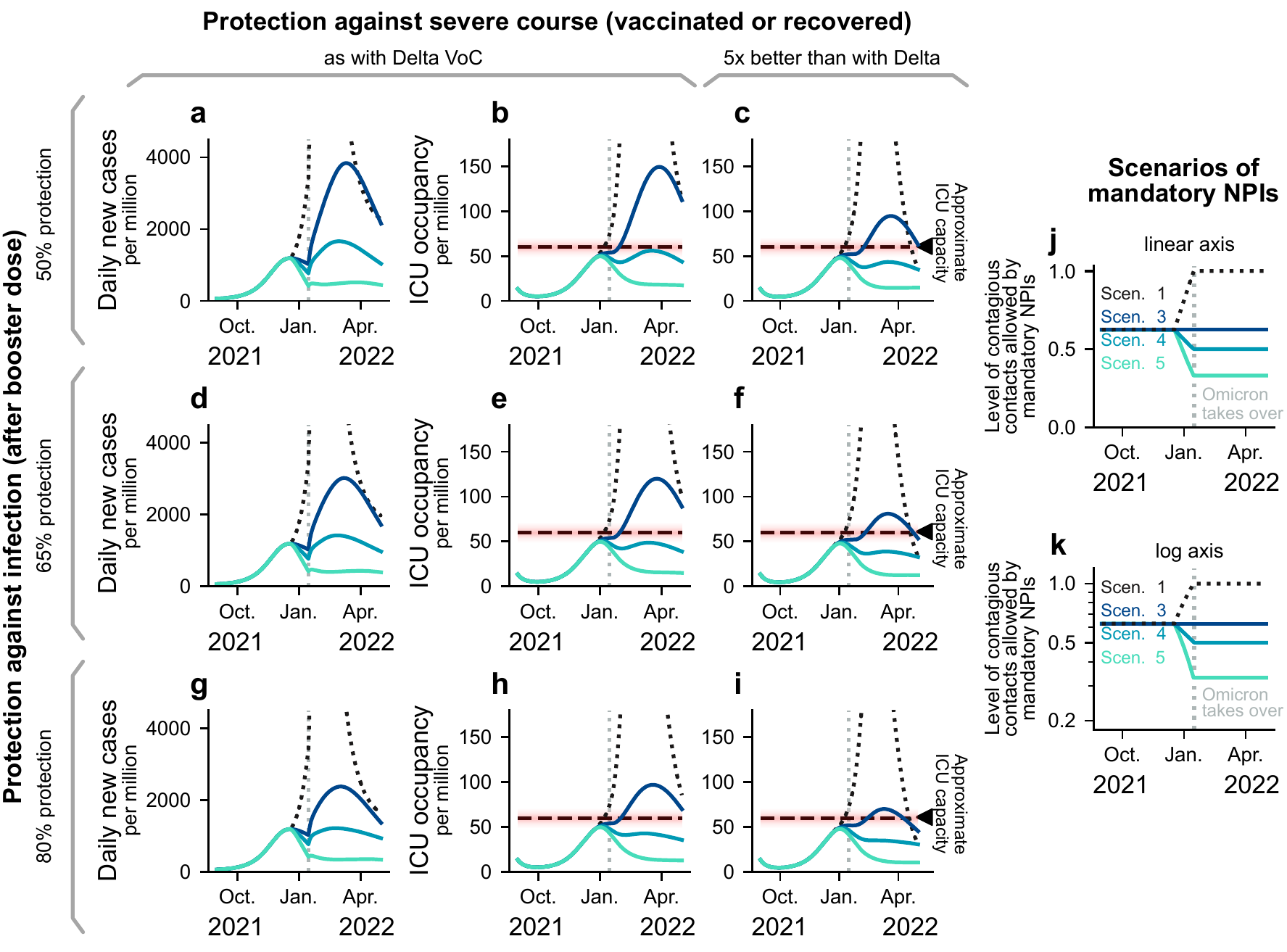}
     \caption{%
         \textbf{Development of the pandemic under the emergence of the Omicron VOC.} Assuming a full replacement of Delta by the Omicron VOC on 15th of January 2022, we model three different possibilities for vaccine-protection against infection, and two levels of long-lasting vaccine- or post-infection protection against severe course (\textbf{a--i}). In colour, we display four scenarios that are derived from the previously studied ones (\textbf{j, k}). All scenarios share moderate mandatory NPIs until mid December 2021, where we evaluate different possibilities for policy adaptation to mitigate the spread of the Omicron VOC. \textbf{a, b, d, e, g, h:} Case numbers and ICU occupancy while assuming that a protection against hospitalisation (once infected despite previous immunisation) is similar to the protection against Delta. \textbf{c, d, i:} ICU occupancy while assuming a protection against hospitalisation (once infected and after previous immunisation) five times better than the protection regarding Delta.
         }
     \label{fig:Figure_6}
\end{figure}

Given this evidence, we analyse the impacts of a potential full replacement of the dominant Delta VOC by the Omicron VOC by 15th of January, 2022. We incorporate the protection against infection by booster doses. As example scenario, we start with Scenario 3 (moderate mandatory NPIs), as it resembles a typical development in Europe. We then analyse four different possible reactions to the Omicron VOC, i.e., starting to switch from Scenario 3 to Scenarios 1, 3, 4, or 5 before it takes over (Fig.~\ref{fig:Figure_6}a). We evaluate three possibilities regarding the booster vaccine-protection against infection, 50\%, 65\%, and 80\% (relative to the protection granted for Delta). This is consistent with available evidence suggesting Omicron's immune escape to reduce vaccine effectiveness against symptomatic disease to about 73\% for freshly mRNA-boosted individuals \cite{cele2021omicron_escape}. Furthermore, we explore two possibilities of severity of infections after previous immunisation: Either efficacy against severe course remains the same as with Delta, both for the immunised and immune-naive persons (Fig.~\ref{fig:Figure_6}b, e, h), or protection is five times better for the immunised (Fig.~\ref{fig:Figure_6}c, f, i). 

As expected, the enhanced transmissibility resulting from the partial escape of the Omicron VOC breaks the decreasing trend in case numbers observed for Scenarios 3, 4, and 5 from the moment where the replacement takes place (Fig.~\ref{fig:Figure_6}a, d, g). This results in a substantial surge in daily new cases in all scenarios except for Scenario 5 (most restrictive). Regarding ICU occupancy, our results depend strongly on the assumed protection against infection by recent vaccination or boosters. When the protection against infection granted by recently administered vaccines is above 50\%, both Scenarios 4 (which has a more strict testing policy and further reduced contacts compared to Scenario 3) and 5 (in addition, group sizes in school are reduced) yield optimistic results for ICU occupancy. If Omicron infections lead to much less severe course of the disease for immunised or convalescent individuals, then even scenario 3 can avoid severely overfilling intensive care units. We have represented Scenario 1 (lifting all mandatory NPIs) with dashed lines, as it yields unrealistic results: Stricter NPIs would probably be reinstated if ICU occupancy becomes too high. The scenarios end in April, where we expect that an updated booster vaccine is developed and distributed. In that phase, lifting restrictions at the pace of vaccination and aiming for low case numbers would maximise freedom while minimising mortality and morbidity \cite{bauer2021relaxing, contreras2020low,oliu2021sars,czypionka2022benefits}.

\clearpage

\section*{Discussion}

Modelling the interplay of human behaviour and disease spread is one of the grand challenges of infectious disease modelling. While not being the first to model behavioural adaptation \cite{funk2010modelling,verelst2016behavioural,weston2018infection,bedson2021review,buonomo2020effects,epstein2021triple}, we incorporate data-driven insights into our modelling framework, inspiring the explicit functional dependency between risk and health-protective behaviour as well as vaccine hesitancy in the context of the COVID-19 pandemic. Thereby, we can incorporate self-regulation mechanisms into our scenario analysis, which best qualitatively describe what is to be expected in the future or in the event of the emergence of novel SARS-CoV-2 VOCs, such as the Omicron variant. We hence take a further step towards more empirically-grounded mathematical models. 

Within our framework, a smooth transition to SARS-CoV-2 endemicity requires, besides a working and accepted vaccine, two ingredients. First, mandatory NPI levels should be high enough to prevent a surge in case numbers so fast that individuals could not react on time to prevent overwhelming ICUs. Second, mandatory NPIs should leave enough room so that individuals can effectively adopt voluntary preventive actions as a response to an increased perception of risk. Hence, governments must guarantee that the decision to, e.g., attend non-essential face-to-face activities that could be carried out remotely remains in the individual's hands. Under such circumstances, voluntary actions can dampen the wave and prevent overwhelming ICUs (scenarios 2 and 3, Fig.~\ref{fig:Figure_5}). Otherwise, irresponsible or overprotective measures would result in a wave that could surpass the healthcare capacity in the short term or when lifting all measures (Scenarios 1, 4, and 5, Figs.~\ref{fig:Figure_4} and~\ref{fig:Figure_5}). In any case, people's awareness about the danger of a disease should ideally be driven by trust in scientific and governmental bodies instead of by the current burden to the healthcare system. Hence, it is crucial during a disease outbreak to engage in extensive, expert-guided, and audience-tailored risk communication \cite{priesemann2021towards} and to prevent the spread of mis- and dis-information that could damage general trust \cite{cinelli2020covid,banerjee2021covid}.

Despite the empirical basis of our approach, the functional shape of the feedback mechanisms remains one of the main uncertainties in our model. The voluntary adoption of health-protective measures was inspired by survey data \cite{betsch2020monitoring}, and is thus bound to its limitations. Additionally, as ICU capacity was never extremely overwhelmed in Germany in the time frame of the COSMO survey, the study does not provide information on how people would act at very high levels of ICU occupancy; in principle, such emergency situations would trigger even stronger reactions in the population, and certainly also a change in NPI stringency (which we assumed to be constant throughout). Furthermore, when extrapolating our results to other countries, one should consider cultural differences or varying levels of trust in governmental bodies. Therefore, more empirical research to inform model assumptions and parameters remains crucial.

Vaccine uptake and coverage are critical parameters that determine mortality and morbidity levels. In line with what has been observed in high-income countries, we assume that vaccination rates are mostly limited by vaccine hesitancy instead of vaccine stocks or logistics. In that way, we can deal with emergent VOCs (as Omicron) with a healthy combination of mandatory NPIs aiming for low-case numbers while a working vaccine is developed and coverage is insufficient \cite{bauer2021relaxing,contreras2020low} and by letting individuals decide on their own when the roll-out is complete. However, the core problem remains latent; wealthy countries concentrate resources while some countries cannot afford enough vaccines to protect even their population at risk \cite{contreras2022rethinking}. As the latter countries are forced into accepting high-case numbers in order to keep their economies running, there are increased risks of breeding variants that could escape natural or vaccine-elicited protection \cite{thompson2021incidence_and_escape}. Therefore, vaccine policy planning from an international perspective is critical for a smooth transition to SARS-CoV-2 endemicity.

Modelling the introduction and spread of different SARS-CoV-2 variants in a population is challenging. At the very least, modelling these dynamics would require having separate compartments for all the disease states of all circulating variants, disproportionally increasing the complexity of our model. In our approach, we take advantage of the extensive immune escape of the Omicron VOC to natural and vaccine-elicited neutralisation \cite{cele2021omicron_escape,lyngse2021sars,viana2021rapid,ferguson2021growth,torjesenn2021omicron_doublingtime}, and assume that the replacement of Delta VOC occures very quickly (i.e. basically instantaneously) in mid-January. This simplification is not too distant from reality; replacement of Delta and other predominant sublineages for Omicron took only a few weeks in several countries \cite{owidcoronavirus}. For the spread of Omicron, we use the same basic reproduction number as for Delta but instead consider most individuals previously immunised to have lost protection against infection, i.e. they are moved to the susceptible pool (Methods for details). Thereby, we can capture the explosive spread of Omicron VOC without increasing the base transmissibility. We furthermore include that those people having received a booster vaccine maintain some protection against infection with Omicron, which, however, also wanes. These assumptions are consistent with a large Danish cohort of households, where the secondary attack rate among unvaccinated was slightly higher for Delta infections than for Omicron \cite{lyngse2021sars}, and with extensive experimental and observational studies \cite{viana2021controlling,cele2021omicron_escape,perez2021dominican,lulu2021}. Despite the approximation we did for the transition to the Omicron variant, the mid- and long- term dynamics of the Omicron VOC should be reflected well. 

In our work, the level of mandatory NPIs dictates the minimum and maximum level of voluntary health-protective behaviour that individuals may adapt. For each scenario, we assume one specific, static level of mandatory NPIs, which best resembles real-world observations on compulsory measures aiming to reduce the probability of contagion (i.e., mask-wearing mandates, immunity passports, meeting restrictions, among others) and testing policy (as described in Methods). However, this static level can lead to unrealistically high waves of incidence and ICU occupancy, which (i) have not been seen so far and (ii) would undoubtedly trigger the implementation of additional restrictions to prevent a major collapse in the health system. Nonetheless, we decided to incorporate this static mandatory NPI level because it illustrates a worst-case trajectory of each scenario. Besides, due to \textit{pandemic fatigue} \cite{petherick2021worldwide}, we would expect the effectiveness of interventions and thus the imposed change in health-protective behaviour in the different mandatory NPI scenarios to decay over time.

In summary, the way governments approach a pandemic situation when vaccines are available will shape long-term transmission dynamics by influencing the magnitude of information-behaviour feedback loops. We show that the latter play a major role during the transition from epidemicity to endemicity. Thus most importantly, the challenge for authorities is to find ways to engage individuals with vaccination programs and health-protective behaviour without requiring high case numbers for that. Here, clear communication and trust continues to be essential \cite{iftekhar2021}.

\section*{Methods}

\subsection*{Model overview}

We use an age-stratified compartmental model with compartments for susceptible-exposed-infected-recovered (SEIR) as well as for fatalities (D), receiving treatment in an ICU (ICU), and vaccination (first time and booster vaccines) (V) (Fig.~S1). We also include waning immunity and seasonality effects (Fig.~\ref{fig:Figure_4},\ref{fig:Figure_5}b). To account for behavioural change induced by perceived risk of infection, we include a feedback loop between ICU occupancy, voluntary health-protective behaviour and willingness to receive vaccination (Fig.~\ref{fig:Figure_2} and Supplementary Information). Explicitly, we assume that increases in ICU occupancy i)~decrease the contact rates among the population and thus the spreading rate of COVID-19 \cite{imbriano2021online,perrotta2021behaviours,druckman2021affective,betsch2020monitoring}, and ii)~increase vaccine acceptance among hesitant individuals \cite{salali2021effective,betsch2020monitoring}. For the first feedback loop (voluntary health protective behaviour), we assume that individuals adapt their contacts in different contexts depending on the risk they have perceived recently. The level of potentially contagious contacts is multiplied by a factor $k$ that decreases with ICU occupancy between the minimum and maximum allowed by current mandatory NPIs (Fig.~\ref{fig:Figure_2}c). Regarding the second feedback loop (related to vaccine uptake), we assume that a fraction of the population will always accept a vaccination offer, despite current ICU occupancy.  From this minimum onward, vaccination willingness monotonically increases with ICU occupancy and saturates towards a maximum, accounting for a fraction of the population that will never accept the vaccine (Fig.~\ref{fig:Figure_2}e). This means that we assume that there is a fraction in the population that is certainly not able or willing to be vaccinated. Given a fraction of people willing to be vaccinated, we determine the speed of the vaccination programme using a linearly increasing function (Fig.~\ref{fig:Figure_2}f). 
We model these two feedback loops to act on different timescales, as individuals can, e.g., decrease the number of contacts and contact intensity on a daily basis, while getting vaccinated takes longer. To capture this, we explicitly include memory kernels accounting for how individuals subjectively weigh events happening on different timescales when forming their perception of risk \cite{zauberman2009discounting}.

\subsection*{Memory on perceived risk}

\label{sec:memoryICU}

We assume that perceived risk regarding the disease depends on information about ICU occupancy that reaches individuals via media or affected social contacts. This perception of risk builds over time; people are not only aware of the occupancy numbers at the present moment but also of those in the recent past. To incorporate this into our model, we calculate the convolution of the ICU occupancy with a Gamma distribution (Fig.~S2, Supplementary Information), effectively "weighting" the ICU occupancy numbers with their recency into a variable of risk perception which we call $\avICU_R$. As a result, ICU occupancy numbers from a few days ago weigh more in people's memory and thus influence voluntary health-protective behaviour at the present moment more than ICU occupancy that lies further in the past. We use this concept of ICU occupancy "with memory" to design the functions of the feedback loops (Fig.~\ref{fig:Figure_2}b, c, e, f).
The effect of the parameters chosen for the Gamma distribution on the model results as well as of all other model parameters is quantified in the sensitivity analysis,  section S4 Supplementary Information. 

\subsection*{NPI- and risk-induced change in health-protective behaviour}\label{sec:selfregulation}

When analysing the joint effect of mandatory NPIs and voluntary measures to mitigate the spread of COVID-19, we find a strong overlap between them; mandatory NPIs limit the range of the measures that individuals could voluntarily take to protect themselves and their loved ones. For example, when large private gatherings are officially forbidden, individuals cannot voluntarily choose not to meet. Additionally, when the engagement of the population in voluntary protective measures is very large, certain mandatory NPIs would not be required. 
We model the combined effect of mandatory NPIs and voluntary adoption of health-protective behaviour as a function $k_{\rm NPI,self}(\avICU_R)$. Using the baseline of mandatory NPIs as an input, this function calculates the level of voluntary preventive action in dependence of the perceived risk $\avICU_R$. To be precise, the value of $k_{\rm NPI,self}(\avICU_R) \in [0,1]$ represents the level to which (potentially contagious) contacts of an average individual are reduced (Fig.~\ref{fig:Figure_2}c), a factor that is multiplied onto the entries of a contact matrix separated by contexts (Fig.~S3, Supplementary Information). For example, adaption of voluntary mask-wearing or a direct reduction of gatherings decreases the level of potentially contagious contacts and, thereby, $k_{\rm NPI,self}(\avICU_R)$. 
Furthermore, we distinguish between contacts made at home, in schools, in workplaces or during communal activities. We weight all the interactions with different $k_{\rm NPI,self}^\nu(\avICU_R)$ with $\nu \in \left\{\text{Households, Schools, Workplaces, Communities}\right\}$ that act on contextual contact matrices $\CMij^\nu$, see Sec.~S1.2 and Fig.~\ref{fig:Figure_1}.

Inspired by the COSMO survey data \cite{betsch2020monitoring} (Fig.~\ref{fig:Figure_2}b), we suggest the following shape for $k_{\rm NPI,self}^\nu(\avICU_R)$:
The level of (potentially) contagious contacts decreases linearly upon increases in the ICU-mediated perception of risk $\avICU_R$ below a threshold $H_R= H_{\rm max}$, from which point on no further reduction is possible (Fig.~\ref{fig:Figure_2}c).
This might represent 
i) a fraction of the population agnostic to measures or unwilling to comply, or
ii) limitations of voluntary preventive action imposed by practical constraints related to the current level of imposed restrictions, for example, having to make contacts in one's own household or having to go to work or school. We implement $k_{\rm NPI,self}^\nu(\avICU_R)$ as a softplus function, having a differentiable transition at $H_{\rm max}$.
Each function (for each scenario) is defined by 3 parameters $H_{\rm max}$, $k_{\rm NPI,self}^\nu(H_R=0)$, and $k_{\rm NPI,self}^\nu(H_R=H_{\rm max})$. $H_{\rm max}=37$ is obtained by the fit to the COSMO data shown in Fig.~\ref{fig:Figure_2} (black line) and used for the two other fits shown in Fig.~\ref{fig:Figure_2} (red and yellow lines) as well as for the behaviour parametrisations for the different scenarios (Fig.~S3, Supplementary Information).

\subsection*{Different mandatory NPI scenarios}
\label{sec:scenarios}

We choose to simulate five different scenarios, each having a different level of overall stringency. In the following we briefly describe the scenarios: 

Scenario 1 ('Freedom day'): All mandatory restrictions are lifted, resulting in a factor of $k_{\rm NPI,self}^\nu(\avICU_R=0) = 1 \hspace{0.1cm} \forall \nu$. However, if ICU occupancy increases, we leave room for individuals' voluntary action based on perceived risk to reduce viral transmission: $k_{\rm NPI,self}^\nu(\avICU_R>0) < 1$. We assume that communal activities and workplaces leave more room for voluntary preventive action than households and schools because of the possibility of working from home, avoiding non-essential gatherings etc. This difference is depicted in Supplementary Fig.~S3.  

Scenario 2 (Moderate NPIs A): Easy-to-follow measures are kept in place and potentially contagious contacts at school are reduced to $k_{\rm NPI,self}^{\rm School}(\avICU_R=0) = 0.5$.

Scenario 3 (Moderate NPIs B): Further measures at work (e.g., home office or testing) reduce $k_{\rm NPI,self}^{\rm Workplaces}(\avICU_R=0) = 0.5$.

Scenario 4 (Moderate NPIs C): Further reduction in potentially contagious school contacts and restrictions affecting communal contacts reduce $k_{\rm NPI,self}^{\rm School}(\avICU_R=0) = 0.25$ and $k_{\rm NPI,self}^{\rm Communities}(\avICU_R=0) = 0.5$. 

Scenario 5 (Strong NPIs): Communal activities are further reduced to $k_{\rm NPI,self}^{\rm Communities}(\avICU_R=0) = 0.2$. 

\tabref{tab:scenarios} lists all values for the different scenarios and contexts of interaction between individuals. 
The reduction of household contacts is assumed to remain the same for all scenarios. Note that, as the stringency of measures increases, room for voluntary adoption of health-protective behaviour usually decreases: To give an example, without mandatory measures the level of contact reduction in communal activities lies in the range $1-0.6$, whereas in a scenario with strong mandatory NPIs it lies in the range $0.2-0.1$. The difference between the two bounds effectively measures the room for voluntary actions (0.4 for freedom day vs 0.1 for strong NPIs). An exception are school contacts in which moderate restriction scenarios (2 and 3) display a wider range of possible voluntary action than the freedom day scenario. As health-protective behaviour among children could be encouraged but not imposed, their adherence to rules constitutes a voluntary act.

\begin{table}[htp]
\caption{\textbf{Different scenarios of mandatory NPIs.} Listed are descriptions of the general measures imposed in each scenario as well as the input parameters to the function $k_{\rm NPI,self}^\nu(H_R)$ that modulates the spread. The parameters act as multiplicative factors onto infection terms in our model, thus high parameter values (close to 1) translate to little reduction in infections and low parameters (close to 0) translate to strong reductions in infections. For each cell, the first parameter translates to a reduction at high ICUs ($k_{\rm NPI,self}^\nu(H_R=H_{\rm max})$) and the second parameter to the corresponding reduction at empty ICUs ($k_{\rm NPI,self}^\nu(H_R=0)$), between which we linearly interpolate (Fig.~S3). }
\label{tab:scenarios}
\centering
\begin{tabular}{ll p{5cm} lllll}\toprule
Sc.  &  Name & Description of measures & $k^{\rm Households}$ & $k^{\rm Schools}$ & $k^{\rm Workplaces}$ & $k^{\rm Communities}$   \\\midrule
1         &  'freedom day' & no mandatory measures
& 0.8-1 & 0.8-1 & 0.6-1  & 0.6-1\\
2         &  moderate NPIs A& increased stringency affecting risk of transmission in schools
& 0.8-1 & 0.25-0.5 & 0.5-0.9  & 0.5-0.9\\
3         &  moderate NPIs B& mild NPIs + reduction of transmission at workplaces 
&0.8-1 & 0.25-0.5 & 0.25-0.5  & 0.5-0.9\\
4         &  moderate NPIs C& moderate NPIs + enforcement of restrictions in communal activities
& 0.8-1 & 0.1-0.25 & 0.25-0.5  & 0.25-0.5\\
5         &  strong NPIs & strong NPIs + further restrictions wherever possible 
& 0.8-1 & 0.1-0.25 & 0.25-0.5  & 0.1-0.2\\
\bottomrule     
\end{tabular}%
\end{table}

\subsection*{Modelling the introduction and spread of the Omicron VOC}
\label{sec:omicronmethods}

Modelling the introduction and spread of the Omicron VOC requires modifications to the model compartments, transition rates, and parameters. In particular, these modifications allow us to explore the effects of Omicron's i) extensive immune escape and ii) potential reduced risk for severe course of the disease. We implemented the introduction of Omicron VOC as a total replacement of the previously dominating Delta VOC on Jan 15, 2022. At that moment, we rearrange the distribution of individuals between the "waned" and "immune" compartments, increase the rate of waning immunity to account for Omicron's immune escape, and reduce the probability of having a severe course. Explicitly, before the introduction of the Omicron VOC, the immune population is tracked in additional pseudo-compartments $V^o,\,R^o,\,R^{v,o}$ with a faster waning rate. In that way, there are always less individuals in $V^o,\, R^o,\, R^{v,o}$ than in $V,R,\,R^v$. At the time of variant replacement, $V-V^o,\, R-R^o,\, R^v-R^{v,o}$ individuals are moved from the vaccinated and recovered compartments to the respective waned compartments; individuals previously protected against Delta would now be susceptible to Omicron. We model booster-vaccination protection against infection following a leaky scheme, thus boostered individuals have a probability of $\eta$ of being entirely protected. With probability $1-\eta$, individuals remain in their current compartment but are tracked as if the vaccine had worked successfully.

\section*{Acknowledgements} 
Open Access publication has been enabled by the Max-Planck-Society. Authors with affiliation “1” acknowledge support from the Max-Planck-Society. SC acknowledges support from the Centre for Biotechnology and Bioengineering—CeBiB (PIA project FB0001, ANID, Chile). SB and SM were financially supported by the German Federal Ministry of Education and Research (BMBF) as part of the Network University Medicine (NUM), project egePan, funding code: 01KX2021. AC recieved funding from the Digital Society research program funded by the Ministry of Culture and Science of the German State of North Rhine-Westphalia. MK acknowledges support from the Netherlands Organisation for Health Research and Development (ZonMw), funding code: 91216062, and from the European Union’s Horizon 2020 research and innovation program under grant agreement No 101003480 (Project CORESMA). KN acknowledges support by the German Federal Ministry of Education and Research (BMBF) for the MODUS-COVID project, funding code: 01KX2022A.
We thank Prof. Dr. Cornelia Betsch, her colleagues and her team for publicly sharing survey data of the COVID-19 Snapshot Monitoring (COSMO) study, which enabled a quantitative approach otherwise not possible.

\section*{Author Contributions}

Conceptualisation: PD, JW, SC, SB, JD, VP\\
Methodology: PD, JW, SC, SB, SBM, ENI, VP\\
Software: PD, JW \\
Formal analysis: PD, JW, SC, SB\\	
Writing - Original Draft: all authors \\	
Writing - Review \& Editing: all authors\\
Visualisation: PD, JW, SC\\	
Supervision: SC, MK, MM, KN, ACV, VP\\	

\section*{Conflict of Interest Statement}

The authors declare that the research was conducted in the absence of any commercial or financial relationships that could be construed as a potential conflict of interest.
\section*{Author Contributions}
Conceptualisation: PD, JW, SC, SB, JD, VP\\
Methodology: PD, JW, SC, SB, SBM, ENI, VP\\
Software: PD, JW \\
Formal analysis: PD, JW, SC, SB\\	
Writing - Original Draft: all authors \\	
Writing - Review \& Editing: all authors\\
Visualisation: PD, JW, SC\\	
Supervision: SC, MK, MM, KN, ACV, VP\\	

\section*{Data Availability Statement}

\section*{Data availability}
Source code for data generation and analysis is available online on GitHub \url{https://github.com/Priesemann-Group/covid19_infoXpand_feedbackloop}.


\newpage

\renewcommand{\thefigure}{S\arabic{figure}}
\renewcommand{\figurename}{Supplementary~Figure}
\setcounter{figure}{0}
\renewcommand{\thetable}{S\arabic{table}}

\setcounter{table}{0}
\renewcommand{\theequation}{\arabic{equation}}
\setcounter{equation}{0}
\renewcommand{\thesection}{S\arabic{section}}
\setcounter{section}{0}

\section*{Supplementary Information}
\section{Model}

\begin{figure}[!ht]
    \centering
    \includegraphics[width=6in]{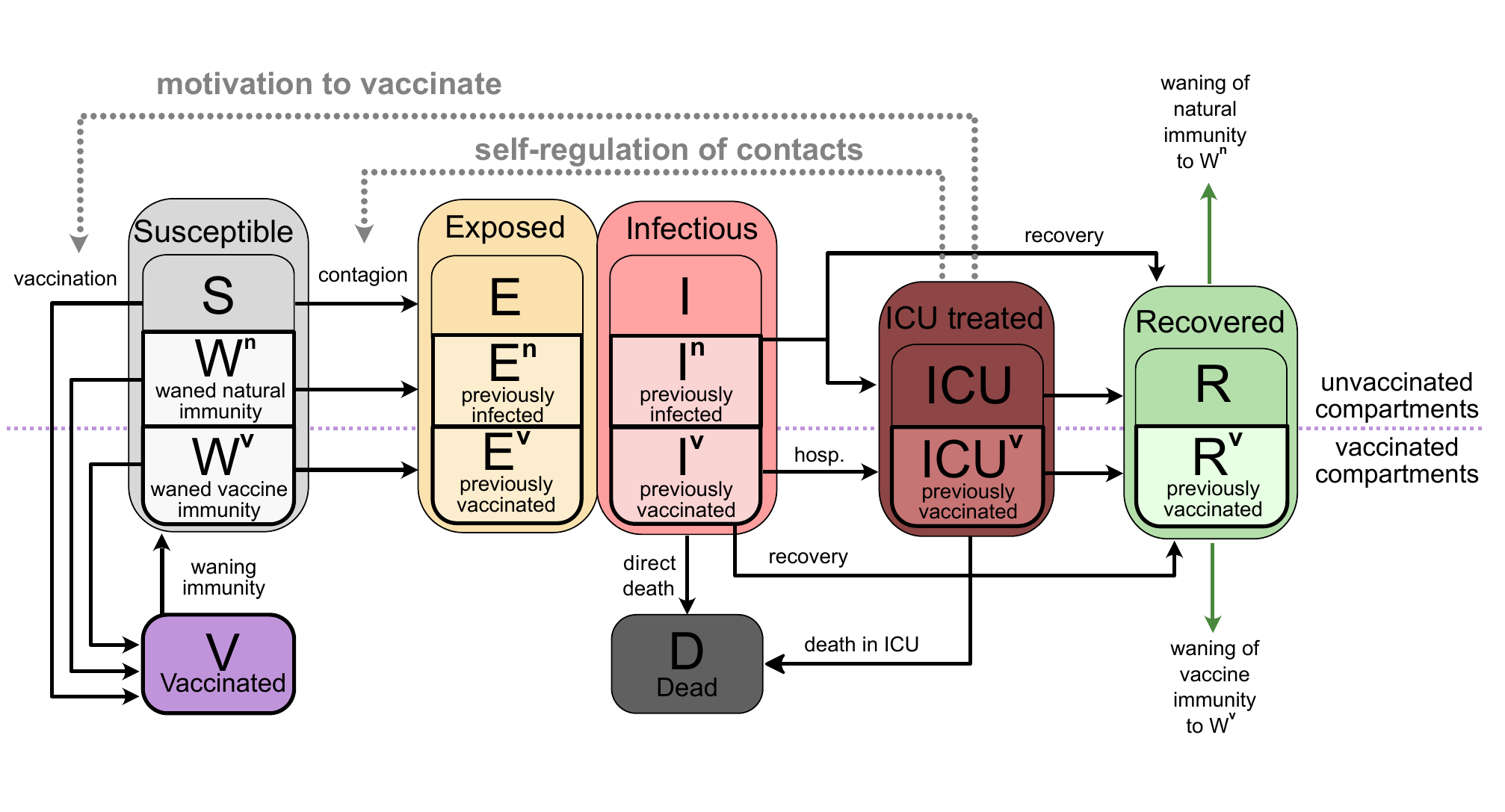}
    \caption{%
        \textbf{Age-stratified SEIRD-ICU compartmental model with vaccination and feedback loops for the interplay between information and disease spread.} Besides considering relevant compartments to capture COVID-19 dynamics, we explicitly incorporate mechanisms of voluntary preventive action through behavioural changes in response to information and individual perception of risks. We incorporate two mechanisms of voluntary action: (i) individuals can voluntarily adapt their immediate health-protective behaviour, adapting it according to their possibilities and the risk they perceive, and (ii) adapt their willingness regarding vaccination, being likelier to accept vaccine offers when feeling at risk for prolonged periods. Transition rates and other variables are listed in Tables~S3 and~S5, but omitted in the figure for clarity purposes.
        }
    \label{fig:Figure_S1}
\end{figure}

We model the spreading dynamics of SARS-CoV-2 by a deterministic age-stratified compartmental model. Our model incorporates disease spreading dynamics (SEIRD), intensive care unit stays (ICU), the roll-out of a single-dose equivalent vaccine and boosters thereof (V), the protection from which wanes over time, and the interplay between risk perception and disease spread through the self-regulation of voluntary health-protective behaviour. We assume that health-protective behaviour is modulated by the perception of risk. When perceiving risks, humans tend to weigh more recent developments more heavily as well as put more weight on developments in the timescale relevant for the decision to be made (i.e., shorter timescales for immediate actions and longer ones for one-time decisions with sustained consequences) \cite{zauberman2009discounting}. Explicitly, if perceiving increased risk, individuals can (i) adapt their level of potentially contagious contacts they have and (ii) adapt their willingness towards seeking vaccination. 
For a graphical representation of the dynamics see~\figref{fig:Figure_S1}. 

In our model, susceptible ($S$) individuals can acquire the virus from infected individuals and subsequently progress to the exposed ($S\rightarrow E$) and, after the latent period, to the infectious ($E\rightarrow I$) compartment. 
Vaccinated and recovered individuals can be infected after their immunity has waned. Alternatively, our model can be interpreted such that waning immunity increases the probability of breakthrough infections. Individuals whose natural or vaccine-induced immunity has waned are modelled via two compartments ($W^n$ and $W^v$, respectively), which feature no protection against infection but against a severe course of the disease, i.e., have reduced probabilities of requiring intensive care or dying. If infected, they transit to different exposed ($\EBn, \EBv$) and infectious ($\IBn, \IBv$) compartments so that vaccinated and unvaccinated individuals are separated. 

The infectious compartments have three different possible transitions: i) direct recovery ($I, \IBn \rightarrow R$ and $\IBv \rightarrow R^v$) with rate $\gamma$, ii) admission to ICU ($I, \IBn \rightarrow \ICU$ and $\IBv \rightarrow \ICU^v$) with rate $\delta$ (reduced by a factor $(1-\kappa)$ for $\IBn,\IBv$) or iii) direct death ($I, \IBn, \IBv \rightarrow D$) with rate $\theta$ (reduced by a factor $(1-\kappa)$ for $\IBn, \IBv$). We assume the recovery and death rate in ICU to be independent of immunisation status. That way, individuals receiving ICU treatment either recover at a rate $\gamma_{\ICU}$ ($\ICU\rightarrow R$ and $\ICU^v \rightarrow R^v$) or die at a rate $\theta_\ICU$ ($\ICU, \ICU^v \rightarrow D$). Note that the probability to get admitted to an ICU is reduced for infected individuals with waned immunity. However, their death rate in ICU is equal to that of those infected for the first time. We use two ICU compartments to separate the vaccinated from the unvaccinated compartments to keep track of individuals who can still receive a vaccine after recovering. 

Each compartment is split into sub-compartments for the age groups that interact with each other following the contact matrix described in Sec.~\ref{sec:spreadingdynamics}. Full age-structured model equations are presented in Sec.~\ref{sec:modelequations}. Apart from the transmission-relevant interactions, the effect of having different age groups is incorporated into our vaccine feedback (described in Sec.~\ref{sec:agevaccineuptake}) as well as in the transition rates between compartments (described in Sec.~\ref{sec: agetransitionrates}).

\begin{figure}[!ht]
    \centering
    \includegraphics[width=6in]{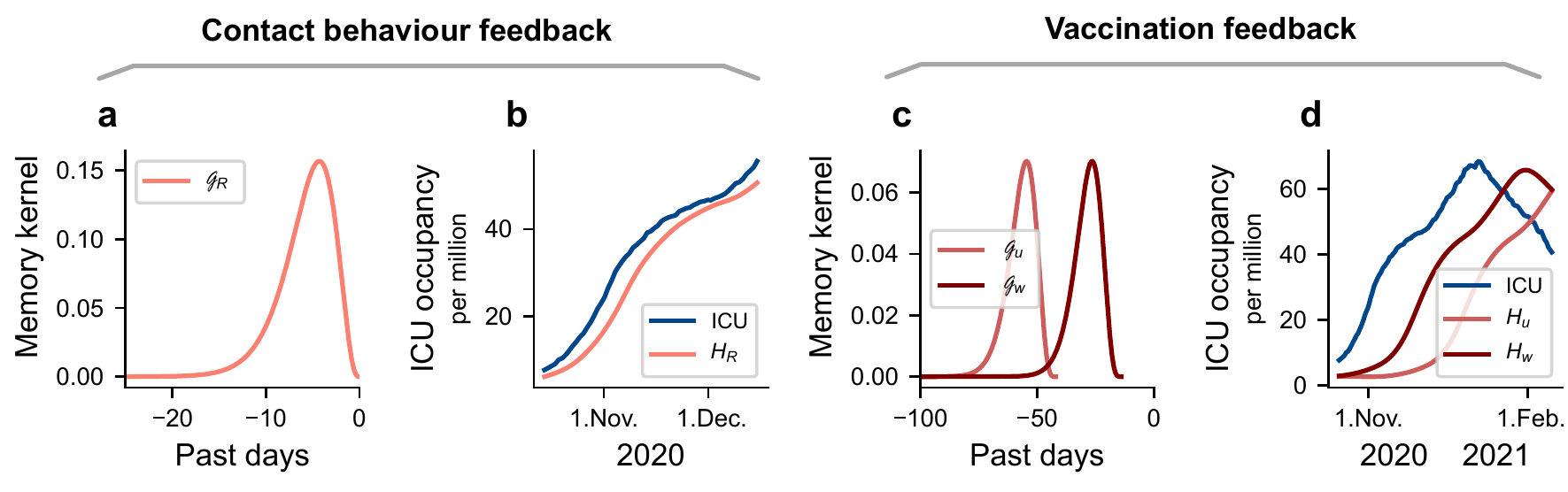}
    \caption{%
        \textbf{Modelling the relationship between perceived risk and pandemic developments.} Based on the information that individuals receive on the recent developments of the pandemic (e.g., ICU occupancy), they form their perception of risk. The way individuals perceive these temporal trends is biased towards recent developments, prioritising them over past developments for their decisions \cite{zauberman2009discounting}. Furthermore, we assume a delay in individuals' reactions to ICU occupancy because of (i) delays inherent to the information spreading dynamics, and (ii) need for recurrent stimuli and various sources for accepting new information. Therefore, we convolve the ICU occupancy time series with a Gamma delay kernel (\textbf{a}), which captures both the delay related to information delivery and the subjective perception of time described above. Vaccine dynamics require a further delay related to the time required to build up immunity: Individuals whose immunity takes effect at a certain time made their decision and got vaccinated some time ago. The length of this delay depends on whether it is a first time or booster vaccination (\textbf{c}). \textbf{b,d:} Once convoluted with ICU occupancy (German example shown here), we obtain a measure for perceived risk $H_R$, $H_u$ and $H_w$, respectively, for the voluntary adaption of immediate health-protective behaviour, first-time vaccination, and booster vaccination. In comparison with the actual ICU development, the variables $H_*$ are smoother and delayed in time, representing non-instantaneous decisions based on individuals' perception of the recent ICU occupancy.}
    \label{fig:Figure_S2}
\end{figure}

\subsection{Memory kernel}
\label{sec:memorykernel}

In this section we specify the memory kernel that measures how risk perception builds on past development of the ICU occupancy. These memory kernels (Fig.~S2) relate to two processes occurring on different timescales. Voluntary adaption of health-protective behaviour depends on the perceived risk in the recent past, $H_R(t)$, defined as:
\begin{equation}
    H_R(t) := \ICU^{\rm tot} \ast \mathcal{G}_{p_R,b_R} = \int_{-\infty}^{t} \text{dt}'\, \ICU^{\rm tot}(t')\,\mathcal{G}_{p_R, b_R}(-t'+t)\,.
\end{equation}
$\ICU^{\rm tot}(t)$ is the sum of all patients in ICU treatment at time $t$: $\ICU^{\rm tot}(t) = \sum_i \ICU_i(t)+\ICU_i^v(t)$. The arguments of the Gamma distribution $\mathcal{G}_{p_R, b_R}$ are set to $p_R=0.7$ (shape) and $b_R=4$ (rate), resulting in a curve that peaks at around four days in the past (Fig.~S2a). Depending on $H_R(t)$, individuals reduce their potentially contagious contacts in different contexts by a weighting factor $k_{\rm NPI,self}$ (Fig.~2, main text) within thresholds determined by current mandatory NPIs (Fig.~S3). See Sec.~\ref{sec:sensitivitynalysis} for a sensitivity analysis on parameter choices. 

Time memory for vaccination willingness is assumed to work in the same way, but with different Gamma distributions, for two reasons. Firstly, there is a delay $\tau_u$ or $\tau_w$ between the decision to be vaccinated and the onset of immunity. Secondly, vaccination willingness is assumed to depend more strongly on past ICU occupancy compared to more immediate health-protective behaviour. Combined, it translates into a Gamma distribution $\mathcal{G}_{p_{\rm vac}, b_{\rm vac}}$ that is shifted in time and is flatter (Fig.~S2c), which is characterised by the parameters $\tau_u$, $\tau_w$, $p_{\rm vac}=0.4$ and $b_{\rm vac}=6$:

\begin{equation}
    H_{u,w}(t) := \ICU^{\rm tot} \ast \mathcal{G}_{p_{\rm vac}, b_{\rm vac}} = \int_{-\infty}^{t} \text{dt}'\, \ICU^{\rm tot}(t')\,\mathcal{G}_{p_{\rm vac}, b_{\rm vac}}(-t'-\tau_{u,w}+t).
\end{equation}

The subscripts $u$ and $w$ indicate first and booster doses, respectively.
Booster doses are usually only a single dose so $\tau_w$ is just the delay between administration of the dose and onset of immunity, which we assume to be 2 weeks. The parameter $\tau_u$ is larger than $\tau_w$ because we include the delay of around 6 weeks for most vaccines that need two doses. For the initial conditions of $\avICU_R$ and $\avICU_{u,w}$, $\ICU$ and $\ICU^v$ are set to a constant $\ICU(t<0)=\ICU(t=0)$ (same for $\ICU^v$) in the past.

\subsection{Spreading dynamics}
\label{sec:spreadingdynamics}

In our model, the spreading dynamics are governed by the sizes of the infectious compartments $I, \IBn, \IBv$ and the compartments $S, W^n, W^v$, from which a transition to an infected state is possible. We include the effects of (i) mandatory non-pharmaceutical interventions (NPIs), (ii) individuals voluntarily adapting their health-protective behaviour based on perceived risk, and (iii) seasonality. Each is represented by a factor $k$ that acts as a multiplicative reduction or increase on the spreading dynamics. 

Seasonality is described by the factor $\kseason$ (see \autoref{eq:seasonality} below). Mandatory NPIs and individuals' voluntary preventive actions are represented by $k_{\rm NPI,self}(\avICU_R)$. It does not factorise into single contributions of mandatory NPIs and voluntary preventive action because we assume the level of NPIs and voluntary behaviour to be coupled.

We introduce an infection term $\sum_j \CM \INF_j$ that governs the spreading between age groups $i$ and $j$. The term is present in all differential equations that include transmissions, i.e., the transitions

\begin{align}
     \begin{array}{ll} \dis S_i \rightarrow E_i & \text{non vaccinated, non infected}\\                       
    \dis W^n_i \rightarrow \EBn_i & \text{waned infected (unvaccinated)} \\
    \dis W^v_i \rightarrow \EBv_i &  \text{waned vaccinated (potentially infected previously)} \end{array} .
    \label{eq: infections1}
\end{align}

include a term proportional to $\sum_j \CM \INF_j$, which is subtracted from the susceptible and waned and added to the exposed states. 

$\CMij$ is the overall contact matrix, which we describe below, and $\INF_i$ is a term describing the infectiousness of age group $i$. We define it as 

\begin{equation}
    \INF_i := \beta \cdot \kseason \cdot \frac{\Ieff_i}{M_i} \hspace{0.5cm} \text{with} \hspace{0.5cm} \Ieff_i := \left(I_i+\IBn_i+\IBv_i+ \Psi M_i\right)\,.
\end{equation}

$\beta$ is the spreading rate, $\Ieff$ is the effective size of the infectious compartments, $M_i$ is the total population size of age group $i$, and $\Psi$ is an external influx of infections, which we assume to be distributed equally over the population, e.g., being the largest for the largest age group. 

\begin{figure}[!ht]
    \centering
    \includegraphics[width=6in]{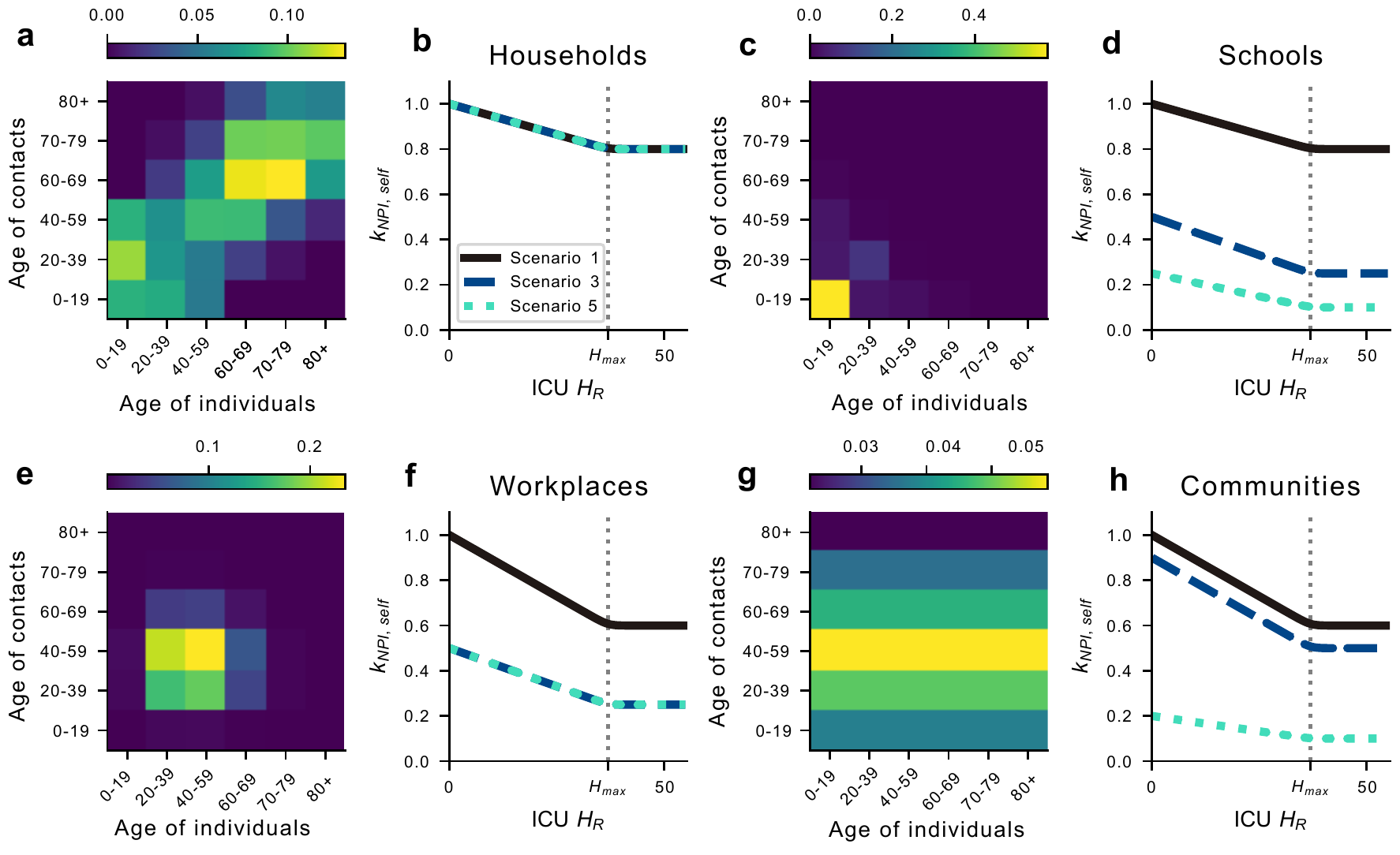}
    \caption{%
        \textbf{The mechanism of the reduction of potentially contagious contacts.} The contact matrix for interactions within households, schools, workplaces and communal activities (\textbf{a,c,e,g}) and the ICU-occupancy-dependent reduction $k_{\rm NPI,self}^\nu(H_R)$ (\textbf{b,d,f,h}) for scenarios 1,3, and 5. Each matrix entry is multiplied by the value of $k_{\rm NPI,self}^\nu(H_R)$ (\textbf{b,d,f,h}), which decreases linearly with perceived ICU occupancy $H_R$ up to the point $H_{max}=37$ where no further reduction is taken as motivated by Fig.~2, main text.}
    \label{fig:matrices}
\end{figure}

The coupling between age groups is represented by a pre-COVID-19 contact matrix $\CMij$. This matrix represents the static, non-ICU-dependent contact behaviour of the different age groups (age group $i$ potentially infecting age group $j$). It can be interpreted as the sum of various layers of contextual contacts (work-, school-, community-, and household-related contacts) \cite{mistry2021inferring}. For a graphical representation of the contextual layers, see~Fig.~1, main text, and \figref{fig:matrices}. Depending on the context, some of these contacts can be voluntarily reduced according to individuals' perception of risk. Hence, we use each of the contextual layers of the matrix $\CMij^\nu$ separately and weigh each layer with reduction factors $k_{\rm NPI,self}^\nu(\avICU_R)$. We use $H_R$ as an effective measure of the ICU occupancy that reflects the population's perceived risk (see \autoref{sec:memorykernel}). 
Finally, we normalise the overall contact matrix $\CMij$ by its spectral radius when its values are not reduced because of mandatory NPIs or voluntary protective behaviour, i.e., at $k_{\rm NPI, self}=1$ and $\avICU_R=0$. That way, the largest eigenvalue of the contact matrix $\CMij = \sum_\nu \CMij^\nu$ equals one in the absence of mandatory NPIs and voluntary measures. 

The resulting infection term present in all transmission-related differential equations for age group $i$ is thus
\begin{equation}
    \sum_j \CM \INF_j = \beta \cdot \kseason \sum_j \left( \sum_\nu \CM^\nu \cdot k_{\rm NPI,self}^\nu \right) \frac{\Ieff_j}{M_j}\,,
\end{equation}
with $j$ counting age groups and $\nu$ counting layers of the contact matrix. Having a normalised contact matrix $\CMij$, we can approximate the seasonal reproduction number $R_{0,\rm seasonal}(t)$, which is defined as the largest eigenvalue of the next generation matrix \cite{diekmann2010construction}, at $\avICU_R=0$. By assuming that $\delta, \theta \ll \gamma$, we get 

\begin{equation}
    R_{0,\rm seasonal}(t) \approx k_{\rm seasonality}(t)\frac{\beta}{\gamma}\,, \label{eq:R0approx}
\end{equation}

with $\gamma = \sum_i \gamma_i M_i$. Postulating that $R_0=5$ at $k_{\rm seasonality}=1$, we can use this formula to calculate the spreading rate $\beta$.
Note that this only holds true if seasons are long compared to the duration of an infection. With the latent period being $\frac{1}{\rho} = 4$ days and the duration in the infected compartment approximately $\frac{1}{\gamma} = 10$ days, the duration of an infection is roughly two weeks which is shorter than the time scale over which seasonality varies significantly.  

We incorporate the effect of seasonality $\kseason$ as a time-dependent sinusoidal modulation factor, as proposed in \cite{Gavenciak2021seasonality}:

\begin{equation}
    \kseason =  1 + \mu \cos\left(2\pi\frac{t+d_0-d_\mu}{360}\right),
    \label{eq:seasonality}
\end{equation} 

where $\mu$ is the sensitivity to seasonality, $d_0$ the starting day of the simulation, and  $d_\mu$ the day with the highest effect on seasonality. We set $d_\mu=0$, corresponding to January 1st. For simplicity, we assume that one month has 30 days and a full year, thus, 360 days. This approximation does not affect the results in the observed time horizon.

\subsection{Contact matrices}
\label{sec:matrices}

In our model, individuals can adapt the level of contagious contacts based on their perception of risk. Explicitly, we consider the contact matrices for the German population reported in \cite{mistry2021inferring}, which differentiate between four different contexts (Households, Schools, Workplaces, and Communities). These matrices are represented in Fig.~S3a, c, e, g. Then, depending on the perception of risk, the scenario of mandatory NPIs, and how much freedom these allow for in different contexts, we calculate a weighting factor $k_{\rm min} \leq k_{\rm NPI,self}^\nu(H) \leq k_{\rm max}$ that multiplies each matrix (Fig.~S3b, d, f, h). Scenario-dependent threshold values for the weighting factors are reported in Table~1 and explained in the Methods Section, main text.

The contact matrix for the Community context is equally distributed, meaning that each individual ("x-axis") has the same probability of being infected by any contact ("y-axis"), independent of age. Because the age groups are different in size, a horizontal pattern emerges; it is likelier to be infected by an individual part of a larger age group. 

Although straightforward to understand, the household layer of contacts applied to our mean-field model may lead to unrealistic results in some situations. For example, consider an ideal full lockdown policy where any transmissions between households were perfectly eliminated. Obviously, in such a theoretical scenario, the pandemic would quickly end as infected individuals would not transmit the virus any further than to contacts within their household. However, under a mean-field compartmental model, the distinction between people in one's household and another household cannot be made, which would lead to a viral spread even under such a scenario. To solve this issue, the factor $k_{\rm NPI,self}^{\text{Households}}(H)$ is scaled by a factor which is the average of the other reductions: $\frac{1}{3}\sum_{\nu}k_{\rm NPI,self}^\nu(H)\,, \nu \in\{\text{Schools, Workplaces, Communities}\}$. In that way, eliminating all contacts in contexts aside from households should end the pandemic.

\subsection{Vaccination effects and waning immunity} \label{sec:vaccinationeffects}

Our model includes the effect of vaccination, where vaccines are for simplicity administered with a single-dosage delivery scheme. Vaccinated individuals cannot be infected while being in the vaccinated compartment, but will proceed to the waned immunity compartment $W^v$ at a rate $\Omega$ \cite{thomas2021safety,puranik2021durability}. The same applies to recovered individuals, who also lose their post-infection immunity at rate $\Omega$ \cite{turner2021sars}. Hence, people transition from compartment $R_i$ to compartment $W_i^n$ and from $R_i^v$ and $V_i$ to $W_i^v$ at rate $\Omega$.

We assume the emptying of the immune compartments to be exponential with rate $\Omega$ or, equivalently, with half-life period $T_{1/2}=\text{ln}(2)/\Omega$. In other words, we assume that after $T_{1/2}$, half of the immune individuals have completely waned immunity and the other half is still fully immune. Within the mean-field approximation, this corresponds to all individuals in the immune compartments having halfway waned immunity after $T_{1/2}$. This time, when the effectiveness against infection $\eta$ reduces to 50\%, equals to about 5 months according to empirical data (for vaccination) \cite{tartof2021effectiveness}. Hence, the waning immunity rate is given by
\begin{equation}\label{eq:Omega}
    \Omega = \frac{\text{ln}(2)}{T_{1/2}} \approx \frac{\text{ln}(2)}{5\cdot30\, \rm days} \approx \frac{1}{225}\frac{1}{\rm days}.
\end{equation}

As soon as individuals enter one of the waned compartments they can be infected with the same probability as individuals never infected or vaccinated before. However, we assume that robustness against a severe course of the disease remains high \cite{tartof2021effectiveness,chemaitelly2021waning,pegu2021durability,naaber2021dynamics} which leads to a reduction of $\left(1-\kappa\right)$ to the probability of requiring treatment in ICUs or dying directly. The parameter $\kappa$ is estimated using $\kappa_\mathrm{obs}$, which denotes the full protection against hospitalisation as in observed studies. The parameter $\kappa$ used in our model is lower than $\kappa_\mathrm{obs}$ because it is the effectiveness against hospitalisation once an individual is already infected. We estimate it via 

\begin{equation}
    (1-\eta)(1-\kappa)=(1-\kappa_\mathrm{obs})  \label{eq: kappa}
\end{equation}
with $\eta$ being the vaccine effectiveness against an infection. According to \cite{tartof2021effectiveness} it holds that $\eta= 0.5$ and $\kappa_\mathrm{obs}=0.9$ (both after five months). Thus, we estimate $\kappa \approx 0.8$ and approximate it to be independent of the time after vaccination.

\subsection{Vaccine uptake}

The age group dependent vaccine uptake is described by two different functions: one for susceptible individuals ($\phi_i$) and one for individuals whose immunity has waned ($\varphi_i$). The core idea is to vaccinate only if willingness for vaccine uptake is larger than the fraction of already vaccinated; if the fraction of individuals who are willing to be vaccinated with a first dose ($u^{\rm willing}$) is larger than the fraction of already vaccinated ($u^{\rm current}$), vaccinations are carried out at a rate proportional to the difference of the two.

Willingness to be vaccinated depends on the past development of the ICU occupancy numbers. $u^{\rm willing}$ can shift between a minimum and a maximum value ($u^{\rm base}$ and $u^{\rm max}= 1-\chi_u$), representing the general observed acceptance for the first dose and people who are strictly opposed to vaccines or cannot be immunised because of age or other preconditions (making up $\chi_u$), respectively. The sensitivity constant $\alpha_u$ determines how sensitive to ICU occupancy the vaccine hesitancy is  (see Sec.~\ref{sec:assessmentAlphas}). 
The willingness to receive the first dose of the vaccine is then described by 

\begin{align}
     u^{\rm willing}_i & = u^{\rm base}_i + \left(u^{\rm max}_i-u^{\rm base}_i\right)\left(1-\exp\left(-\alpha_u \avICU_u \right)\right)\,.
\end{align}

Hence, $u^{\rm willing}_i$ is a fraction for each age group $i$ between zero and one and the total number of people willing to be vaccinated in each age group $i$ is thus $u^{\rm willing}_i M_i$.
For the differences in the parameters $u^{\rm base}_i$ and $u_{i,\rm max}$ between age groups, see Sec. \ref{sec:agevaccineuptake} and for a graphical example representation of $u^{\rm willing}$ see Fig.2e, main text.

The function that determines the rate at which first time vaccines are administered is denoted by $\phi$. It determines the transition away from $S_i$ and $W^n_i$, is age group dependent, and is described via a softplus function:

\begin{equation}
    \phi_i(\avICU_u) = \frac{1}{t_u} \cdot \frac{S_i+W^n_i}{M_i(1 - u^{\rm current}_i)} \cdot \epsilon \ln \left( \exp \left( 1+ \frac{1}{\epsilon} \left( u^{\rm willing}_i(\avICU_u)-u^{\rm current}_i \right) \right) \right)\,,
    \label{eq: phi}
\end{equation}

where $\epsilon$ is a curvature parameter. Multiplying by $\frac{S_i+W^n_i}{M_i(1 - u^{\rm current}_i)}$ ensures that we only vaccinate if people are actually present in $S$ or $W^n$. Dividing by $t_u$ smoothens the transition between the state of vaccinating and not vaccinating, the physical explanation being that people require time (of the order of $t_u$ days) to organise a vaccine, which reduces the vaccination rate after crossing the threshold. $t_u$ is assumed to be constant here. However, when there is a lot of demand for vaccine uptake, $t_u$ is likely larger in reality due to administrative and logistical problems. For the implementation of $\phi_i$ into the model equations, see Sec.~\ref{sec:modelequations}. In the term $\frac{dS_i}{dt}$ we multiply $\phi$ by $\frac{S_i}{S_i+W^n_i}$ and in the term $\frac{dW^n_i}{dt}$ we multiply by $\frac{W^n_i}{S_i+W^n_i}$, effectively splitting up the vaccinations among the two groups. 

The administration of booster doses works in a similar way. First, we define a function for the age group dependent willingness to accept a booster dose:

\begin{align}
    w^{\rm willing}_i & = w^{\rm base}_i + \left(w^{\rm max}_i -w^{\rm base}_i\right)\left(1-\exp\left(-\alpha_w \avICU_w \right)\right)\,.
\end{align}

The function for booster doses $\varphi$ can then be written as

\begin{equation}
    \varphi_i(\avICU_w) = \frac{1}{t_w} \cdot \frac{W^v_i}{M_i(u^{\rm current}_i - w^{\rm current}_i)} \cdot \epsilon \ln \left( \exp \left( 1+ \frac{1}{\epsilon} \left( w^{\rm willing}_i(\avICU_w)u^{\rm current}_i - w^{\rm current}_i \right) \right) \right)\,,
    \label{eq: varphi}
\end{equation}

We only vaccinate if willingness among those who received a first dose is larger than the fraction of already boostered people, i.e. $u^{\rm current}_i$ is the upper limit for $w^{\rm current}_i$. 

\subsubsection{Assessment of sensitivity to ICU occupancy for vaccination dynamics}
\label{sec:assessmentAlphas}

In our model, we assume the willingness in the total population to be vaccinated for the first time to range between threshold values $u^{\rm base}$ and $u^{\rm max}$. The difference $u^{\rm max}- u^{\rm base}$ is the fraction of people that, initially hesitant, decide to accept the vaccine offer based on their perception of risk. In order to estimate how sensitive this group is to risk perception in the form of awareness about the ICU occupancy, we proceed as follows. If we estimate the ICU occupancy at which half of the people belonging to this initially hesitant group accepts a vaccination, we can calculate the sensitivity parameter $\alpha_u$: Let $H_{1/2}$ be this ICU occupancy. We then have to solve 

\begin{equation}
    u^{\rm base}+\frac{1}{2}\left(u^{\rm max}-u^{\rm base}\right) \overset{!}{=} u^{\rm base}+\left(u^{\rm max}-u^{\rm base}\right)\left(1-\text{exp}\left(-\alpha_u H_{1/2}\right)\right),
\end{equation}

which reduces to 

\begin{equation}
    \alpha_u = \frac{\text{log}2}{H_{1/2}}. \label{eq:assessmentalpha}
\end{equation}

We assume $H_{\rm max}$, i.e., the threshold at which no further adaption of health-protective behaviour occurs, as a first estimate for $H_{1/2}$ to obtain an approximate value for the sensitivity as $\alpha_u = \frac{\text{log}2}{H_{\rm max}} = \frac{\text{log}2}{37} \approx 0.02$. The quantified effect that this parameter has on the results is explored in Sec.~\ref{sec:sensitivitynalysis}. 

\subsubsection{Tracking vaccinated individuals}

Transition rates between the susceptible ($S_i$) and waned ($W^n_i, W^v_i$) compartments due to vaccination depend on the difference between willingness to be vaccinated and the fraction of currently vaccinated. Thus, it is necessary to keep track of how many people have received a first and booster dose, respectively. This is modelled by integrating over the vaccination rates. It translates into two additional differential equations:

\begin{equation}
    \frac{d}{dt}u^{\rm current}_i = \phi_i(\avICU_u) \hspace{0.5cm} \text{and} \hspace{0.5cm}  \frac{d}{dt}w^{\rm current}_i = \varphi_i(\avICU_w)\,,
\end{equation}

where $u^{\rm current} $ and $w^{\rm current} $ are the fraction of people who received a first and booster dose, respectively.
The initial conditions for $u^{\rm current}_i$ and $w^{\rm current}_i$ are the total reported numbers of administered vaccine doses \cite{owidcoronavirus}.

\subsection{Exploring vaccination rate and ICU occupancy trends in different European countries}
\label{sec:vaccinesothercountries}

\begin{figure}[!ht]
\hspace*{-1cm}
    \centering
    \includegraphics[width=7in]{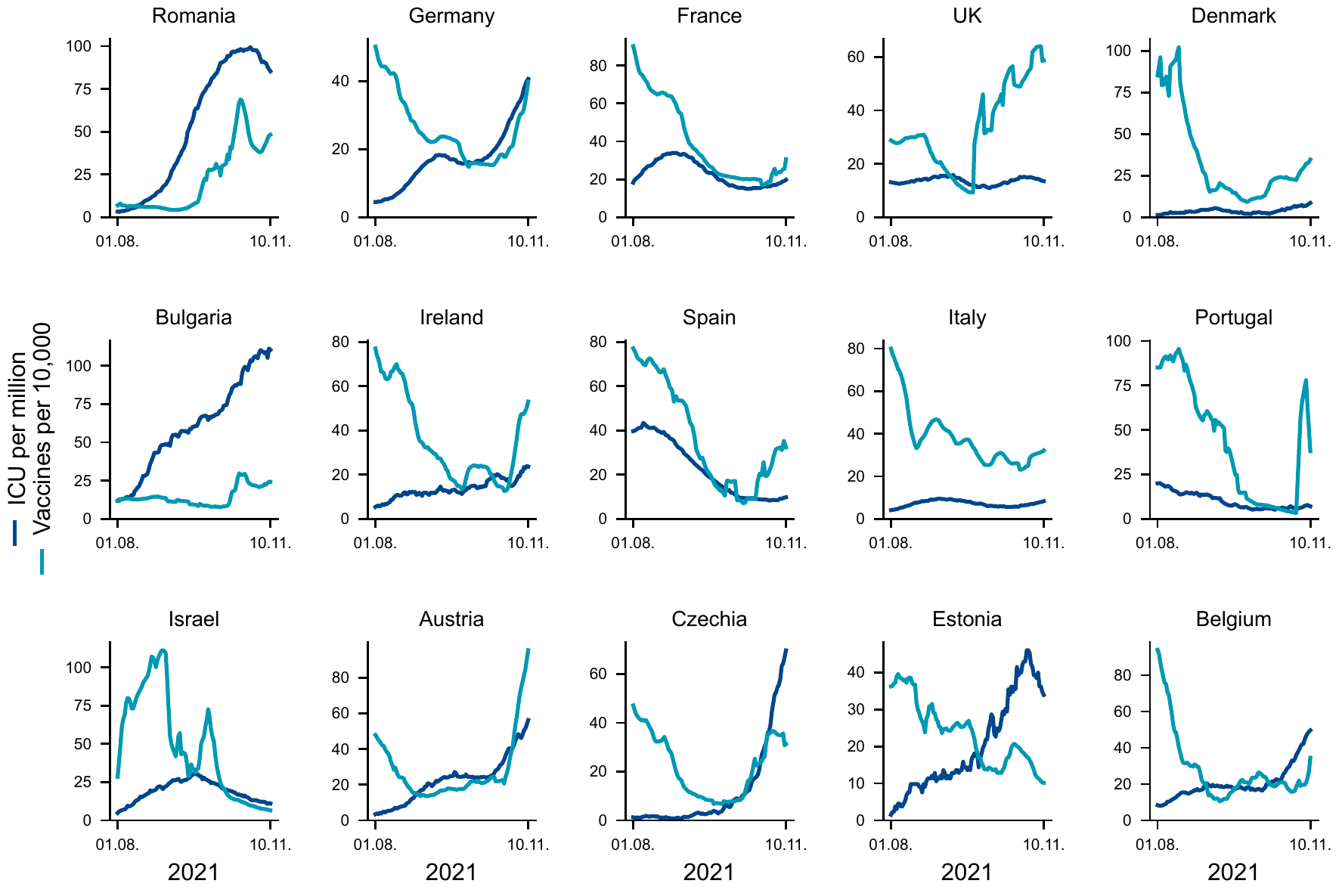}
    \caption{%
        \textbf{Vaccination rate and ICU occupancy trends across selected countries.} ICU occupancy per million inhabitants and daily vaccinations per 10,000 inhabitants for several European countries and Israel. Booster doses and first time doses are added together.}
    \label{fig:vaccinescountries}
\end{figure}

The main assumption underlying the vaccination feedback is that vaccination willingness follows ICU occupancy. In the case of Romania this relation is evident (\autoref{fig:vaccinescountries}): Approaching winter 2021, case numbers and ICU occupancy had a steep rise, arguably due to insufficient immunity among the population, as vaccine coverage was under $30\%$ \cite{owidcoronavirus}. Under such circumstances, there was a lot of "room for improvement" within the unvaccinated population not strictly opposed to vaccines, which led to a steep surge in administered doses (\figref{fig:vaccinescountries}). Note that there might also be other underlying causes for increased vaccine uptake: For example, imposing restrictions only onto unvaccinated might motivate vaccine uptake. While this is a governmental choice not considered in our model, such enforcements usually follow high levels of ICU occupancy and are thus indirectly accounted for. 

In countries other than Romania, the trend is less visible (\figref{fig:vaccinescountries}). Several countries show an increase in vaccine uptake in October 2021; however, it is unclear whether this is mainly motivated by voluntary behaviour following an increase in ICU occupancy. Concurrently, requests for launching country-wide booster campaigns were on the rise, which might have been the leading cause of increased vaccine uptake. However, whether the causes are voluntary behaviour or institutional recommendations regarding vaccinations, both follow perceived risk (on individual level vs governmental level) and lead to the same effect: increased ICU occupancy leads to increased vaccine uptake. Apart from Estonia and Belgium, we do not observe countries in which a rise in ICU occupancy is not followed by a rise in vaccinations. If the contrary is the case, i.e., vaccines rising despite ICU occupancy staying low, this could be attributed to external motivations and would require further country-specific investigation.

In countries where we observe increasing vaccines following ICU occupancy, we should note that the delay between the two varies a lot. While in Romania and Bulgaria, the delay seems to be of the order of one month, we observe that in Germany, Austria and the Czech Republic, there does not seem to be any relevant delay. Note that the vaccination curve measures daily administered vaccines and not the onset of immunity (which the kernel in our model represents). The effect of the delay incorporated in our model is quantified in the sensitivity analysis \ref{sec:sensitivitynalysis}. 

\section{Age stratification}

\subsection{Age-dependent vaccine uptake}
\label{sec:agevaccineuptake}

Although there are vaccines accredited for children below 12 years in the European Union, we assume that these age groups will have much lower uptakes, affecting our parameters $u^{\rm max}_i$ (maximum vaccine uptake) and $w^{\rm max}_i$ (maximum booster uptake). Furthermore, due to likelier side effects of vaccines for the young, but lesser consequences of an infection, we assume that these parameters as well as the baseline acceptances for vaccines increase with age. Thus, $u^{\rm base}_i$, $w^{\rm base}_i$, $u^{\rm max}_i$ and $w^{\rm max}_i$ become age-dependent. All vaccine-related parameters are listed in \tabref{tab:vaccineagegroups}. Note that $w^{\rm max}_i$ is a fraction of $u^{\rm current}_i$ and not of $M_i$, thus it is no contradiction if $w^{\rm base}_i > u^{\rm base}_i$.

\begin{table}[htp]
\caption{\textbf{Different age groups and age-dependent parameters related to vaccine uptake.} }
\label{tab:vaccineagegroups}
\centering
\begin{tabular}{l  cccccc}\toprule
Group ID &  age group & fraction of population $M_i/M$ & $u^{\rm max}_i$ & $w^{\rm max}_i$ & $u^{\rm base}_i$ & $w^{\rm base}_i$  \\\midrule
1           & 0-19 & 0.18 & 0.35 & 0.76 & 0.2 &0.1\\
2      & 20-39 & 0.25 & 0.9 & 0.8 & 0.5 & 0.25 \\
3  & 40-59 & 0.28& 0.92 & 0.84 & 0.55 & 0.275\\
4         & 60-69 & 0.13  & 0.94 & 0.88  &0.6 & 0.3\\
5         & 70-79& 0.09 & 0.96 & 0.92  &0.65 & 0.325\\
6         & 80+ & 0.07& 0.98 & 0.96& 0.7 & 0.35\\
\bottomrule     
\end{tabular}%
\end{table}

\subsection{Age-dependent transition rates}
\label{sec: agetransitionrates}

Differing disease severity after a SARS-CoV-2 infection for different age groups translates into age-dependent transition rates between our model compartments. More specifically, we include age-dependent parameters for the natural recovery rate $\gamma$, the ICU admission rate $\delta$, the death rate $\theta$ and the recovery as well as death rates from ICU, $\gamma_\ICU$ and $\theta_\ICU$, respectively. \tabref{tab:agegroupstransitions} lists the different parameters as reported in \cite{bauer2021relaxing}.

\begin{table}[htp]
\caption{\textbf{Age-dependent transition parameters related to the ICU-, death- and recovery rates.} All parameters are given in units of days$^{-1}$. }
\label{tab:agegroupstransitions}
\centering
\begin{tabular}{l  cccccc}\toprule
ID &  Age group & \makecell[c]{Recovery rate \\ $\gamma_i\,\left[\SI{}{day^{-1}}\right]$}  & \makecell[c]{ICU adm. rate \\ $\delta_i\,\left[\SI{}{day^{-1}}\right]$}  & \makecell[c]{Death rate \\ $\theta_i\,\left[\SI{}{day^{-1}}\right]$} & \makecell[c]{ICU rec. rate \\ $\gamma_{\ICU,i}\,\left[\SI{}{day^{-1}}\right]$}  & \makecell[c]{ICU death rate \\ $\delta_{\ICU,i}\,\left[\SI{}{day^{-1}}\right]$} \\\midrule
1           & 0-19 &0.09998 & 0.000014 & 0.000002 &0.19444 & 0.00556 \\
2      & 20-39 & 0.09978 & 0.000204 & 0.000014 & 0.19222 & 0.00778 \\
3  & 40-59 & 0.09867 & 0.001217 &0.000111 &0.084745 & 0.006164 \\
4         & 60-69 & 0.09565 & 0.004031 & 0.000317 & 0.081401   &0.009508 \\
5         & 70-79& 0.09314  & 0.005435  & 0.001422 & 0.091355  & 0.019756\\
6         & 80+ &0.08809  & 0.007163 & 0.004749  & 0.084233 & 0.082433\\
\bottomrule     
\end{tabular}%
\end{table}

\section{Model Equations}
\label{sec:modelequations}

The combined contributions of the infection-spreading and vaccination dynamics are represented by the set of equations below. The time evolution of our model is then completely determined by the initial conditions of the system. The first-order transition rates between compartments are given by the probability for an individual to undergo this transition divided by the average transition time, e.g., the recovery rate $\gamma$ is the probability that an individual recovers from the disease divided by the time span of the recovery process. Note that in principle $\gamma$ should be different for the $I$ and $\IB$ compartment, as the probability to recover is larger for individuals previously immunised. We neglect this difference as it is negligible within the margin of error since the probability to recover is close to 1 in both cases. The subscripts $i$ denote the sub-compartments for each age group and $C_{ij}$ the contact matrix that describes the interactions within the age groups. 

\begin{align}
&\Ieff_i & = & \xunderbrace{\left(I_i+\IBn_i+\IBv_i+ \Psi M_i\right)}_{\text{effective incidence}}  & &  && \label{eq:Ieff} \\
&\INF_i & = & \xunderbrace{\beta\,\kseason\, \frac{\Ieff_i}{M_i}}_{\text{effective infection rate}}  & &  && \label{eq:INF} \\
&\CMij  & = & \xunderbrace{ \sum_\nu \CMij^\nu \, k_{\rm NPI,self}^\nu}_{\text{sub-matrices times reductions}} & &  && \label{eq:CM}\\
&\frac{d S_i}{dt} & = & -\xunderbrace{S_i \sum_j \CM \INF_j}_{\text{unvaccinated infections}} &-& \xunderbrace{M_i \phi_i(\avICU_u) \frac{S_i}{S_i+W^n_i}}_{\text{first vaccinations}} & &  \label{eq:dSdt} \\
& \frac{dW^n_i}{dt} & = & -\xunderbrace{W^n_i \sum_j \CM \INF_j}_{\text{waned infections}} &-& \xunderbrace{M_i \phi_i(\avICU_u) \frac{W^n_i}{S_i+W^n_i}}_{\text{first vaccinations}}  &+& \xunderbrace{ \Omega R_i}_{\text{waning natural immunity}} \label{eq:dWndt}\\
& \frac{dW^v_i}{dt} & = & -\xunderbrace{W^v_i \sum_j \CM \INF_j}_{\text{waned infections}} &-& \xunderbrace{M_i u^{\rm current}_i \varphi_i(\avICU_w)}_{\text{booster vaccinations}}  &+& \xunderbrace{\Omega V_i+ \Omega R^v_i}_{\text{waning immunity}} \label{eq:dWvdt}\\
& \frac{dV_i}{dt} & = & & & \xunderbrace{M_i\left(\phi_i(\avICU_u)+u^{\rm current}_i \varphi_i(\avICU_w)\right)}_{\text{vaccinations }} &-&  \xunderbrace{\Omega V_i}_{\text{waning vaccine immunity}} \label{eq:dVdt} \\
& \frac{dE_i}{dt} & = & \xunderbrace{S_i \sum_j \CM \INF_j}_{\text{unvaccinated exposed}}   &-&\xunderbrace{\latRate E_i}_{\text{end of latency}} & &  \label{eq:dEdt}\\
&\frac{d\EBn_i}{dt} & = & \xunderbrace{W^n_i \sum_j \CM \INF_j}_{\text{unvaccinated waned exposed}}  &-&\xunderbrace{ \latRate \EBn_i}_{\text{end of latency}}  & &  \label{eq:dEBndt}\\ 
&\frac{d\EBv_i}{dt} & = & \xunderbrace{W^v_i \sum_j \CM \INF_j}_{\text{vaccinated waned exposed}}  &-&\xunderbrace{ \latRate \EBv_i}_{\text{end of latency}}  & &  \label{eq:dEBvdt}\\ 
&\frac{dI_i}{dt} & = & \xunderbrace{\latRate E_i}_{\text{start of infectiousness}}   &-&\xunderbrace{ \left(\gamma_i + \delta_i + \theta_i\right) I_i}_{\rightarrow\text{recovery, ICU, and death}} & &   \label{eq:dIdt}\\ 
&\frac{d\IBn}{dt} & = & \xunderbrace{\latRate \EBn_i}_{\text{start of infectiousness}}   &-&\xunderbrace{ \left(\gamma_i + (\delta_i+\theta_i)(1-\avProtection)\right) \IBn_i}_{\substack{\rightarrow\text{recovery, ICU (reduced),}\\\text{and death (reduced)}}}  & &\label{eq:dIBndt}\\ 
&\frac{d\IBv_i}{dt} & = & \xunderbrace{\latRate \EBv_i}_{\text{start of infectiousness}}   &-&\xunderbrace{ \left(\gamma_i + (\delta_i+\theta_i)(1-\avProtection)\right) \IBv_i}_{\substack{\rightarrow\text{recovery, ICU (reduced),}\\\text{and death (reduced)}}}  & &\label{eq:dIBvdt}\\ 
&\frac{d\ICU_i}{dt} & = & \xunderbrace{\delta_i \left(I_i+(1-\avProtection)\IBn_i\right) }_{\text{nonvaccinated ICU}}  &-&\xunderbrace{ (\gamma_{\ICU, i}+\theta_{\ICU, i}) \ICU_i}_{\text{recovery or death in ICU}}  & & \label{eq:dICUdt}\\
&\frac{d\ICU^v_i}{dt} & = & \xunderbrace{\delta_i (1-\avProtection)\IBv_i}_{\text{vaccinated ICU}}  &-&\xunderbrace{ (\gamma_{\ICU, i}+\theta_{\ICU, i}) \ICU^v_i}_{\text{recovery or death in ICU}}  & & \label{eq:dICUvdt}\\
&\frac{dD_i}{dt} & = & \xunderbrace{\theta_i \left(I_i + (1-\avProtection)\left(\IBn_i+\IBv_i\right)\right)}_{\text{death without ICU}}   &+&\xunderbrace{ \theta_{\ICU, i} \left(\ICU_i+\ICU^v_i\right)}_{\text{death in ICU}}  & & \label{eq:dDdt}\\
&\frac{dR_i}{dt} & = & \xunderbrace{\gamma_i (I_i+\IBn_i)}_{\text{direct recovery}} & + & \xunderbrace{ \gamma_{\ICU, i}\, \ICU_i}_{\text{ recovery}}   &-&\xunderbrace{ \Omega R_i}_{\substack{\text{waning}\\\text{post-infection immunity}}}  \label{eq:dRdt}\\
&\frac{dR^v_i}{dt} & = & \xunderbrace{\gamma_i \IBv_i}_{\text{direct recovery}} & + & \xunderbrace{\gamma_{\ICU, i} \, \ICU^v_i}_{\text{recovery from ICU}} &-&\xunderbrace{ \Omega R^v_i}_{\substack{\text{waning}\\\text{post-infection immunity}}}  \label{eq:dRvdt}\\
&\frac{du^{\rm current}_i}{dt} & = & \xunderbrace{\phi_i(\avICU_u)}_{\text{current first vaccinations}}  & &  && \label{eq:dIUCdt}\\ 
&\frac{dw^{\rm current}_i}{dt} & = & \xunderbrace{\varphi_i(\avICU_w)}_{\text{current booster vaccinations}}  & &  && \label{eq:dIWCdt}\\ 
\end{align}

\begin{table}[htp]\caption{\textbf{Model parameters (in order of first appearance) related to infection dynamics. $^*:$\cite{Levin2020}\cite{salje2020estimating}\cite{bauer2021relaxing}\cite{Linden2020DAE}} The parameters referring to \autoref{tab:agegroupstransitions} are age-dependent.}
\label{tab:Parametros}\small
\centering
\begin{tabular}{l p{4cm} lll p{2cm}}\toprule
Pa  & Meaning & \makecell[l]{Value \\ (default)}     & Unit  &   Source  \\\midrule
$\gamma$            & Recovery rate            & Tab. \ref{tab:agegroupstransitions}  & \SI{}{day^{-1}}  & \cite{he2020temporal,pan2020time} \\
$\delta$            & Avg. hospitalisation rate ($I\to\ICU$)       & Tab. \ref{tab:agegroupstransitions}  & \SI{}{day^{-1}}  &  $^*$ \\
$\kappa$   & Reduction of hospitalisation rate (given infection) for individuals with waned immunity & 0.8   & \SI{}{-}   & Eq.~\ref{eq: kappa}\\
$\theta$       & Avg. death rate & Tab. \ref{tab:agegroupstransitions}  & \SI{}{day^{-1}}  &  $^*$\\
$\gamma_\ICU$  & Recovery rate from \ICU    & Tab. \ref{tab:agegroupstransitions} & \SI{}{day^{-1}}  &  $^*$ \\
$\theta_\ICU$  & Avg. \ICU death rate   & Tab. \ref{tab:agegroupstransitions} & \SI{}{day^{-1}}  &  $^*$ \\
$\CMij$              & Contact matrix      & \SI{}{-}        & \SI{}{-} & \cite{mistry2021inferring} \\
$\beta$     &  Spreading rate  & 0.5  &  \SI{}{day^{-1}} & Eq.~\ref{eq:R0approx} \\
$\Psi$ & Influx of infections & 1    & \SI{}{people/day}  & Assumed\\
$R_0$               & Basic reproduction number (Delta variant) &   5.0  &  \SI{}{-} &  \cite{liu2021reproductive} \\
$\rho$   & Rate of leaving exposed state                        & 0.25       & \SI{}{day^{-1}}  & \cite{bar2020science, li2020substantial}\\
$\mu$  & Sensitivity to seasonality                         & 0.267    & --  & \cite{Gavenciak2021seasonality} \\
$d_0$    & Day when the time series starts                   & 240    & \SI{}{day}  & Chosen \\
$d_\mu$  & Day with the strongest effect on seasonality     & 0    & \SI{}{day}  & \cite{Gavenciak2021seasonality}\\
$\Omega$  & Waning immunity rate (base) & $\frac{1}{225}$   & \SI{}{day^{-1}}   & \cite{tartof2021effectiveness}, Eq.~\ref{eq:Omega} \\
$\fracImmun$        & Vaccine eff. against transmission 5 months after vaccination   & 0.5    & \SI{}{-}   & \cite{tartof2021effectiveness}\\
$\kappa_\mathrm{obs}$       & Observed vaccine eff. against severe disease 5 months after vaccination & 0.9   & \SI{}{-}        & \cite{tartof2021effectiveness} \\
\bottomrule
\end{tabular}%
\end{table}

\begin{table}[htp]\caption{\textbf{Model parameters (in order of first appearance) related to the behavioural feedback loops.} The range column describes the range of values used in the various scenarios.}
\label{tab:Parametros2}
\centering
\begin{tabular}{l p{7cm} lll p{2cm}}\toprule
Parameter  & Meaning & \makecell[l]{Value \\ (default)}     & Unit  &   Source  \\\midrule
$p_R$, $b_R$ & Shape and rate parameters of the memory kernel for the risk perception relevant to immediate health-protective behaviour, respectively & 0.7, 4.0 & \SI{}{-} & Assumed \\
$\tau_u, \tau_w$    & Memory time of the ICU capacity and delay to immunisation    & 2, 6     & \SI{}{weeks}  & Assumed \\
$p_{\rm vac}$, $b_{\rm vac}$ & Shape and rate parameters of the memory kernel for the risk perception relevant to vaccination, respectively & 0.4. 6.0  & \SI{}{-} & Assumed \\
$k^\nu$  & Weighting factors for the contextual contact matrices & Tab. 1, main text   & \SI{}{-}    &  Assumed\\
$u^{\rm base}, w^{\rm base}$  & Base fractions of vaccine acceptance (first and booster, respectively) & Tab. \ref{tab:vaccineagegroups}         & \SI{}{-}  &  \cite{wouters2021challenges}\\
$\chi_u, \chi_w$ & Fraction of the population refusing vaccine (first and booster, respectively) & Tab. \ref{tab:vaccineagegroups}    & \SI{}{-}  &  \cite{betsch2020monitoring}\\
$\alpha_u, \alpha_w$            & Sensitivity of the population to ICU occupancy   & 0.02    &  \SI{}{people^{-1}} &  Eq.~\ref{eq:assessmentalpha} \\
$\epsilon$ & Curvature parameter for the softplus function describing the vaccination rate & 1  & \SI{}{-} & Chosen\\
$t_u, t_w$ & Organization time for vaccine (first and booster resp.) & 7 & days & Assumed\\
$H_{\rm max} $ & Risk perception above which no further adoption of voluntary health-protective behaviour occurs & 37    & \SI{}{-}        & Fitted to \cite{betsch2020monitoring}\\\bottomrule
\end{tabular}%
\end{table}

\begin{table}[htp]\caption{Model variables.}
\label{tab:Variables}
\hspace*{-1cm}
\centering
\begin{tabular}{l p{4cm} l  p{9cm} }\toprule
Variable & Meaning & Unit & Explanation\\\midrule
$M$               & Population size       & \SI{}{people} & Default value: 1,000,000\\
$S$ & Susceptible compartment & \SI{}{people} & Non-infected people, who may acquire the virus.  \\
$V$ & Vaccinated compartment & \SI{}{people} & Non-infected, vaccinated people. Less likely to be infected or develop severe symptoms  \\
$W^n$ & Waned post-infection immunity compartment & \SI{}{people} & Non-infected people whose post-infection immunity has already waned, thus may acquire the virus.  \\
$W^v$ & Waned vaccine immunity compartment & \SI{}{people} & Non-infected people whose vaccine-induced immunity has already waned, thus may acquire the virus.  \\
$E$ & Nonvaccinated exposed compartment   & \SI{}{people} & Nonvaccinated, non-previously-infected people exposed to the virus. \\
$\EBn$ & Nonvaccinated, waned exposed compartment  & \SI{}{people} & Nonvaccinated, previously-infected people exposed to the virus whose post-infection immunity has waned.\\
$\EBv$ & Vaccinated exposed compartment  & \SI{}{people} & Exposed people with waned vaccine immunity.\\
$I$ & Infectious compartment   & \SI{}{people} & Infectious people from the susceptible compartment $S$. \\
$\IBn$ & Nonvaccinated, waned infectious compartment     & \SI{}{people} & Infectious people from $\EBn$.\\
$\IBv$ & Vaccinated infectious compartment     & \SI{}{people} & Infectious people with waned vaccine-induced immunity.\\
$\ICU$ & Nonvaccinated hospitalised     & \SI{}{people} & Nonvaccinated hospitalised people (from $I$ and $\IBn$) .\\
$\ICU^v$ & Vaccinated hospitalised     & \SI{}{people} & Previously-vaccinated, hospitalised people (from $\IBv$) .\\
$R$ & Unvaccinated Recovered & \SI{}{people} & Unvaccinated recovered people (with or without requiring intensive care).\\
$R^v$ & Vaccinated Recovered    & \SI{}{people} & Vaccinated recovered people (with or without requiring intensive care).\\
$H_*$ & Avg. ICU occupancy   & \SI{}{people} & Auxiliary variable measuring the memory on past ICU occupancy.\\
$u^{\rm current}, w^{\rm current}$ & Vaccinated individuals, independent of the compartment  & \SI{}{-} & Integral over the vaccination rates $\phi, \varphi$. \\
$\kseason$ & Seasonal variation of SARS-CoV-2 transmission &  \SI{}{-} & Eq.~\ref{eq:seasonality}. \\
$k_{\rm NPI, self}$ & Reduction of infections due to mandatory NPIs and voluntary behaviour  & \SI{}{-} & Sec.~\ref{sec:matrices} \\
$\phi(t), \varphi(t)$ & Administration rate of first-time and booster vaccine doses (resp.)  &  \SI{}{doses/day}        &  Eq.~\ref{eq: phi}, \ref{eq: varphi}\\
\bottomrule
\end{tabular}%
\end{table}

\subsection{Initial conditions}
\label{sec:initials}

A primary task for defining the initial conditions is distributing the population size of $M=10^6$ individuals onto our model compartments (\figref{fig:Figure_S1}). In reality, however, there are no well-defined compartments. For example, a person vaccinated a few months ago cannot be classified into either a $V$ or $W^n$ compartment, but is instead in a vaccinated state with reduced vaccine effectiveness. Furthermore, available data on vaccinated or infected individuals is often age-stratified by different age groups or not age-stratified at all. To approach these data challenges, we obtain the initial conditions through the following procedure (\figref{fig:initials}):

\begin{figure}[!ht]
    \centering
    \includegraphics[width=6in]{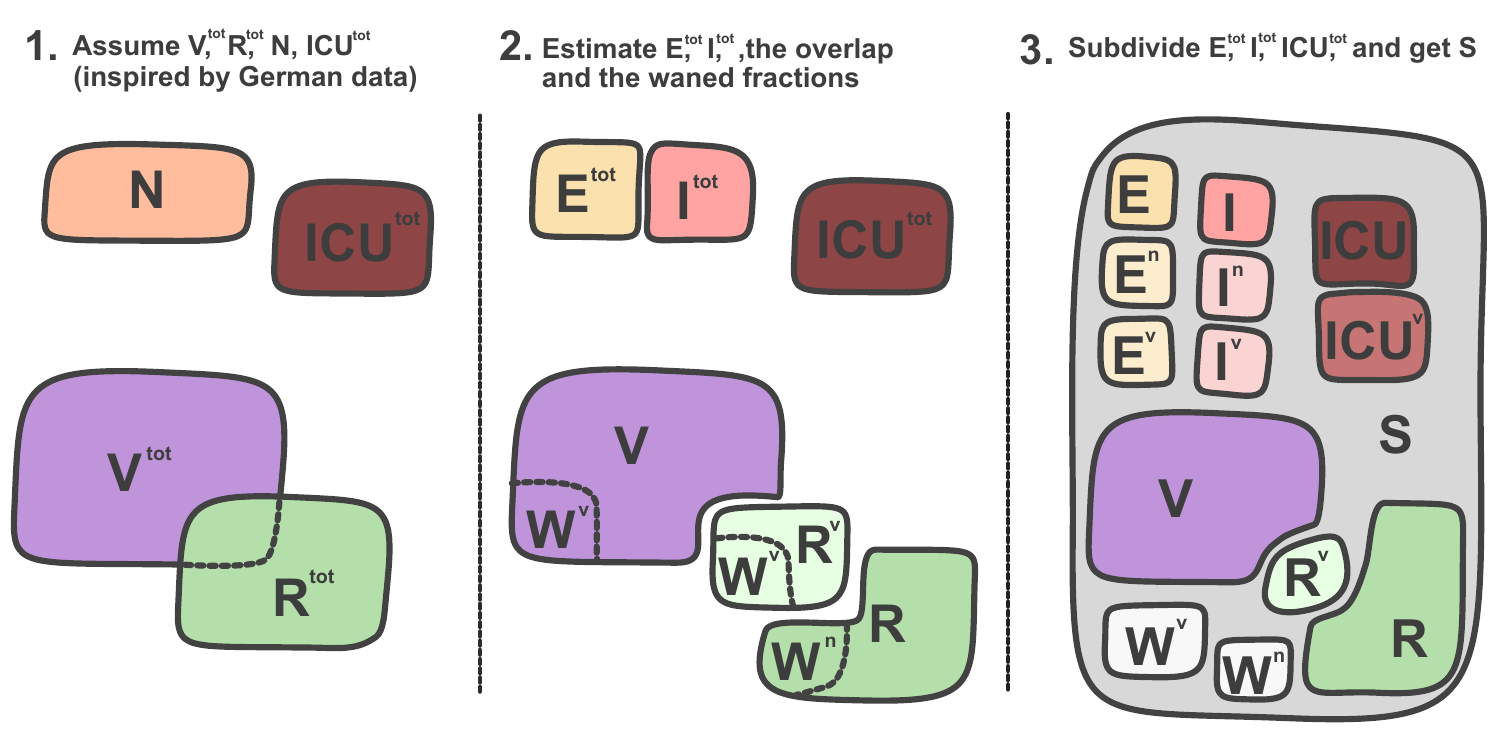}
    \caption{%
        \textbf{The procedure of obtaining initial conditions for the model compartments.} Starting with parts of the population attributed to different states $N$, $\ICU$, $V^{\rm tot}$, and $R^{\rm tot}$, we calculate reasonable values for the initial conditions of all compartments step by step. Compartment sizes in the figure are chosen arbitrarily and do not represent actual size in terms of people.}
    \label{fig:initials}
\end{figure}

We postulate that we want to look at a population that is $60\%$ vaccinated and $20\%$ recovered (including the non reported cases). Let the resulting numbers of people be called $V^{\rm tot}=0.6M$ and $R^{\rm tot}=0.2M$, respectively. These values are inspired by the situation in Germany as of September 1st 2021. Next, we take German data on daily new infections $N$ and ICU occupancy $\ICU^{\rm tot}$ at this point in time \cite{owidcoronavirus, DIVI2020Tagesreport}. These four values will be used to build all the other compartments. First, however, we have to uniformly age-stratify these values. ICU occupancy and the number of new COVID-19 cases can be obtained in an age-stratified way for the case of Germany. For the number of vaccinated and recovered, we assess countries that report age-stratified data, such as Denmark, and distribute the total numbers $V^{\rm tot}$ and $R^{\rm tot}$ onto the various age groups as can be seen in Tab. \ref{tab:initials}. Given the initial values for $V^{\rm tot}_i$, $R^{\rm tot}_i$, $N_i$ and $\ICU^{\rm tot}_i$ for every age group $i$, we calculate the values for all other compartments:

\textbf{Immune compartments separated by vaccination status and previous infection}\\
First, we consider the possibility that individuals were both previously vaccinated and infected. Thus, to avoid overestimating the number of immunised individuals, we estimate the overlap between $V^{\rm tot}_i$ and $R^{\rm tot}_i$: As a first order estimate, we assume that the probability of being vaccinated and having recovered are independent of each other. That way, the probability of being both vaccinated and recovered is given as the product of the two probabilities:

\begin{equation}
    \text{Prob}(x \in V^{\rm tot}_i \land x \in R^{\rm tot}_i) = \text{Prob}(x \in V^{\rm tot}_i ) \cdot \text{Prob}(x \in R^{\rm tot}_i )
\end{equation}

Accordingly, the fraction of vaccinated in the total population for age group $i$, $\frac{V^{\rm tot}_i}{M_i}$, is the same as the fraction of vaccinated in the recovered part of the population, $\frac{R^v_i}{R^{\rm tot}_i}$. Hence, the initial numbers of recovered vaccinated, $R^v_i$, and unvaccinated individuals, $R_i$, are estimated via
\begin{align}
    R^v_i = \frac{V^{\rm tot}_i}{M_i} R^{\rm tot}_i \text{ and } R_i = R^{\rm tot}_i - R^v_i .
\end{align} 
Consequently, we receive the number of vaccinated individuals without previous infection by subtracting the overlap:
\begin{equation}
    V_i = V^{\rm tot}_i - R^v_i.
\end{equation}
This process is illustrated in Fig.~\ref{fig:initials}.

\textbf{Waned compartments separated by immunity status}\\
Next, we consider the fraction of vaccinated and recovered individuals whose immunity has waned (see Tab.~\ref{tab:initials}):
For the recovered, we assume that the time point at which infections took place in the past was age-independent and thus attribute the same fraction of waned natural immunity to all age groups. However, this assumption does not hold for vaccine-induced immunity because older age groups were typically vaccinated at an earlier point in time. We subtract the waned fractions from the compartments $V_i, R_i$ and $R^v_i$, obtaining $W^n_i$ and $W^v_i$.  

\textbf{Susceptible compartment}\\
The susceptible compartment $S_i$ comprises of all individuals not belonging to any of the other compartments. It can be calculated via
\begin{equation}
    S_i = M_i - V_i - R_i - R^v_i - W^n_i - W^v_i - N_i - \ICU^{\rm tot}_i.
\end{equation}

\textbf{Exposed and infectious compartments separated by immunity status}\\
We estimate the initial conditions for the exposed and infected compartments by first estimating $E_i^{\rm tot} = E_i+\EBn_i+\EBv_i$ and $I_i^{\rm tot} = I_i+\IBn_i+\IBv_i$ by

\begin{equation}
    E_i^{\rm tot} = \frac{1}{\rho}N_i \hspace{0.5cm} \text{and} \hspace{0.5cm} I_i^{\rm tot} = \frac{1}{\gamma_i+\delta_i+\theta_i}N_i\,.
\end{equation}

The fractions $\frac{1}{\rho}$ and $\frac{1}{\gamma+\delta+\theta}$ are the average times spent in the exposed and infected compartments, respectively (approximately).

To find out how $E_i^{\rm tot}$ and $I_i^{\rm tot}$ distribute onto their sub-compartments, i.e., for the different immune status and age groups, we look at their origin:  
Because all the infections in $E_i$ originate from $S_i$, the ones in $\EBn_i$ from $W^n_i$ and the ones in $\EBv_i$ from $W^v_i$, we can distribute $E_i^{\rm tot}$ onto the sub-compartments via 

\begin{equation}
    E_i = \frac{S_i}{S_i+W_i^n+W_i^v}E_i^{\rm tot}, \hspace{0.5cm}  \EBn_i = \frac{W_i^n}{S_i+W_i^n+W_i^v}E_i^{\rm tot}, \hspace{0.5cm}  \EBv_i = \frac{W_i^v}{S_i+W_i^n+W_i^v}E_i^{\rm tot}
\end{equation}
and analogously for $I_i^{\rm tot}$. 

\textbf{ICU compartments separated by vaccination status}\\
To determine the distribution of $\ICU^{\rm tot}_i$ onto the compartments $\ICU_i$ and $\ICU_i^v$, we consider that the probability to require ICU care for individuals in the compartments $\IBn_i$ and $\IBv_i$ is reduced by a factor of $(1-\kappa)$. Hence,
\begin{equation}
    \ICU_i^v=\frac{\IBv_i(1-\kappa)}{I_i + (\IBn_i+\IBv_i)(1-\kappa)}\ICU_i^{\rm tot} \quad \text{ and } \quad \ICU_i = \ICU_i^{\rm tot}-\ICU_i^v.
\end{equation}

The initial condition for the dead is set to $D_i=0$, for the currently vaccinated to $u_i^{\rm current}=V_i^{\rm tot}$ and for the currently boostered to $w_i^{\rm current}=0$. For the initial condition of $\avICU_*$, values of past ICU occupancy development are needed. Here, we assume a constant past value of the ICU occupancy at $t\leq d_0$ for both ICU compartments.

\begin{table}[htp]
\caption{\textbf{Initial conditions by age group.}  The total population size in the model is $M=10^6$. The column $\frac{V_i^{\rm tot}+R_i^{\rm tot}-R^v_i}{M_i}$ shows the effective fraction of the population that is immune, which for the entire population is $68\%$ (with $\sum_i R_i^{\rm tot}/M=0.2$ and $\sum_i V_i^{\rm tot}/M=0.6$). Sources: 1: \cite{bauer2021relaxing}, 2: \cite{owidcoronavirus}, 3:\cite{DIVI2020Tagesreport}}
\label{tab:initials}
\centering
\begin{tabular}{l  ccccccccc}\toprule
ID &  age group & $M_i/M$ & $\frac{V_i^{\rm tot}}{M_i}$ & $\frac{R_i^{\rm tot}}{M_i}$ &  $N_i$ & $\ICU_i^{\rm tot}$ & $\frac{W_i^v}{V_i+R^v_i}$ & $\frac{W_i^n}{R_i}$ & $\frac{V_i^{\rm tot}+R_i^{\rm tot}-R^v_i}{M_i}$ \\\midrule
1           & 0-19 & 0.18 & 0.15 & 0.2 & 18.5 & 0.14 & 5\% & 50\% & 0.32\\
2      & 20-39 & 0.25 & 0.56 & 0.2 & 16.8 & 1.24  & 5\% & 50\% & 0.65\\
3  & 40-59 & 0.28& 0.67 & 0.2 & 15.9 & 4.90 & 10\% & 50\% & 0.74\\
4         & 60-69 & 0.13  & 0.77 & 0.2  &6.4 & 3.10 & 20\% & 50\% & 0.82\\
5         & 70-79& 0.09 &0.88 & 0.2  &3.5 & 2.46 & 30\% & 50\% & 0.90\\
6         & 80+ & 0.07& 0.95 & 0.2& 2.3 & 1.62 & 40\% & 50\% &0.96\\ \hline
Source & - & 1 & assumed & assumed & 2 & 3 & assumed & assumed & calculated\\
\bottomrule     
\end{tabular}%
\end{table}

\section{Sensitivity analysis}
\label{sec:sensitivitynalysis}

The results of this model depend on the choices of all parameters involved. While some epidemiological parameters are well understood and quantified at this point in the pandemic, some other parameters of our model remain uncertain, but might have a large impact on the results. 
In this section we analyse the sensitivity of our results to changes in parameters. We vary each parameter independently across its assumed range (see Sec.~\ref{sec:sensitivityparamranges}) and look at how this affects the maximal ICU occupancy observed in the first (winter) and second (spring) waves. We choose a moderate scenario (Scenario 3) for the analysis and look at how the two peaks of ICU occupancy (one in winter, one after lifting restrictions) change in magnitude.

\subsection{Sensitivity to additional parameters}
\label{sec:sensitivitynewparams}

\begin{table}[htp]\caption{\textbf{Additional model parameters introduced in the sensitivity analysis.}}
\label{tab:Parametros_supp}
\centering
\begin{tabular}{l p{7cm} lll p{2cm}}\toprule
Parameter  & Meaning & \makecell[l]{Value \\ (default)}     & Unit  &   Source  \\\midrule
$\sigma$      & Relative viral load of recovered/vaccinated individuals & 1  & \SI{}{-}        & \cite{levine2021viral} \\
$\Omega_n$      & Waning rate of post-infection immunity & $\frac{1}{125} $ & \SI{}{day^{-1}}      & \cite{tartof2021effectiveness} \\
$\Omega_v$      & Waning rate of vaccine immunity & $\frac{1}{125}$  & \SI{}{day^{-1}}       & \cite{tartof2021effectiveness} \\
$\xi$      & Shape of the seasonality function $k_{\rm seasonality}$ & 1  & \SI{}{-}        & \cite{Gavenciak2021seasonality}
\\\bottomrule
\end{tabular}%
\end{table}

For a more precise analysis we introduce new parameters to our model (Tab~\ref{tab:Parametros_supp}). Firstly, we consider the possibility of previously immunised individuals having a reduced viral load and thus being less infectious. This has been reported for vaccinated individuals e.g. in \cite{harris2021impact} for the Alpha variant of SARS-CoV-2, but is unclear for current and future variants. In the model, it can be represented by a change in $\Ieff$, introducing a parameter $\sigma$ for reduced viral load in the infectious compartments $\IBn$ and $\IBv$:

\begin{equation}
    \Ieff_i = \left(I_i+\sigma(\IBn_i+\IBv_i)+ \Psi M_i\right)
\end{equation}

Next, we include the possibility that post-infection and vaccine-induced immunity wane at different rates $\Omega_n$ and $\Omega_v$, respectively. Lastly, we introduce a parameter that affects the shape of $k_{\rm seasonality}$. The transmission of SARS-CoV-2 is strongly reduced in outdoor encounters in comparison to indoor encounters. Thus, winter typically offers more opportunities for viral spread than summer because more activities are performed inside. However, the transition between summer and winter might look different than the standard sinusoidal suggested in Eq.~\ref{eq:seasonality}. In particular, it could be the case that above a certain temperature most activities move outside all at once, resulting in a steeper transition between summer and winter as soon as temperatures allow for it. To model this, we introduce an exponent $\xi\in [0,1]$ that modifies the sinusoidal:

\begin{equation}
    \kseason =  1 + \mu \cdot \sgn\left(\cos\left(t^\star\right)\right) \cdot \left|\cos\left(t^\star\right)\right|^\xi \quad \text{ with } \quad t^\star = 2\pi\frac{t+d_0-d_\mu}{360}\,.
\end{equation}

That way, for $\xi \rightarrow 0 $ the cosine in $k_{\rm seasonality}$ becomes a step function.

\subsection{Parameter ranges}
\label{sec:sensitivityparamranges}

The way we vary parameters differs between age-dependent and non-age-dependent parameters as well as between parameters bound to the $[0,1]$ interval (e.g., $\kappa$) and those belonging to arbitrary intervals. For the age-independent parameters $\kappa, \sigma, \xi \in [0,1]$ we vary them in the range $[0.5,1]$ (for $\kappa$ and $\sigma$) and $[0,1]$ (for $\xi$). For the age-dependent rates with arbitrary range, $\delta_i$, $\gamma_{\ICU, i}$, $\theta_i$, and $\theta_{\ICU, i}$, we consider a range around their default value by a factor of two. For example, for $\delta_i$ we vary across the ranges $[\frac{\delta_i^{\rm default}}{2}, 2\delta_i^{\rm default}]\, \forall i $ at the same time for all age-groups. Figure ~\ref{fig:sensitivity_2} summarises these results.

Parameters related to the memory kernel $p_R, b_R, p_{\rm vac}$, and $b_{\rm vac}$ as well as the sensitivities to vaccine uptake $\alpha_u$ and $\alpha_w$ are also varied around their default value by a factor of two.

For age-dependent parameters related to vaccine uptake $u^{\rm base}_i, w^{\rm base}_i, \chi_{u_i} $, and $\chi_{w_i}$ which are bound to the interval $[0,1]$, we look at their base value multiplied by a factor in the range $[0.8,1.2]$ and vary one parameter for all age groups at the same time. Figure ~\ref{fig:sensitivity_3} summarises these results. 
Parameters $\tau_u,\, \tau_w,\, t_u,\, t_w,\, H_{\rm max}$, and the influx $\Psi$ are varied in a range chosen broad enough such that an effect is observable. 

The average immunity waning times $(\Omega^n)^{-1}$ and $(\Omega^v)^{-1}$ are varied in the range between 4 months and 1 year and the waning rates thus is the range of the inverse values.

\subsection{High impact parameters}
\label{sec:sensitivityhighimpact}

In this section we discuss parameters that have a large impact on the quantitative results when being varied. 

As expected, the waning rate of vaccine-induced immunity $\Omega^v$, leads to much higher waves when increased. The peak of the wave after lifting restrictions is more than doubled for an average waning time of $4.5$ months instead of the $7.5$ months used as default. 

The vaccine efficacy $\kappa$ also plays an important role in the second wave, as by that time, most infections will originate from the waned compartments. 

Naturally, the transition rates to ICU $\delta_i$ have a large impact on the magnitude of the waves. Interestingly, the impact is a lot stronger for the second wave than for the first wave. The reason is that the first wave mainly affects the unvaccinated younger age groups that are less likely to transition to ICU, whereas the second wave affects all age groups similarly. 

One of the main uncertainties in our model is the choice of the sensitivity parameters $\alpha_u$ and $\alpha_w$ that modulate vaccine uptake in dependence of risk perception $H_u$ and $H_w$. Lower values imply a population less reactive to threat, which results in higher waves as can be seen in Fig.~\ref{fig:sensitivity_3}. On the other hand, for large values of $\alpha_u$ and $\alpha_w$, ICU occupancy seems to plateau, not decreasing any further. This suggests a limitation on what voluntary vaccination alone can do to prevent bringing ICUs to capacity limits (given our model assumptions).

\begin{figure}[!ht]
    \centering
    \includegraphics[width=6in]{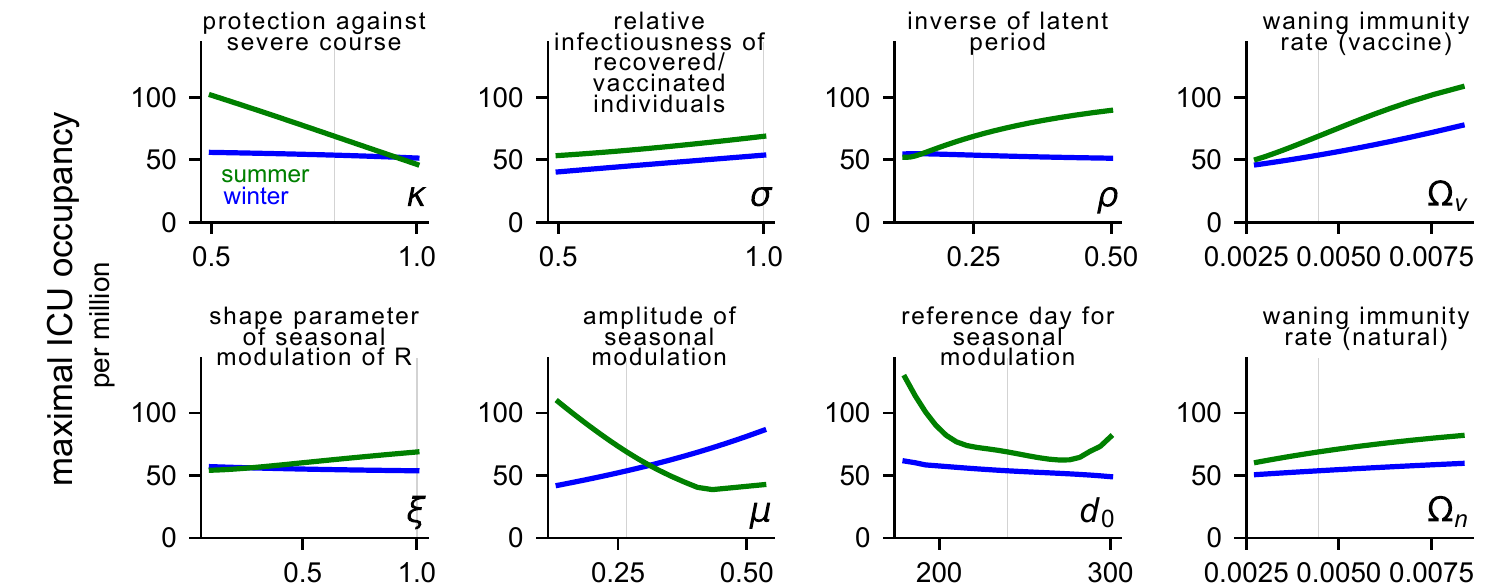}
    \caption{%
        \textbf{Propagation of parameter uncertainties.} Parameters are varied independently across their assumed range. The measured quantity is the maximum ICU occupancy observed in the first wave (blue) and the second wave (green). A vertical line indicates the default value of the parameter. Thus, the points where the green and blue curve intersect the vertical line have the same y-coordinate in all plots. }
    \label{fig:sensitivity_1}
\end{figure}

\begin{figure}[!ht]
    \centering
    \includegraphics[width=6in]{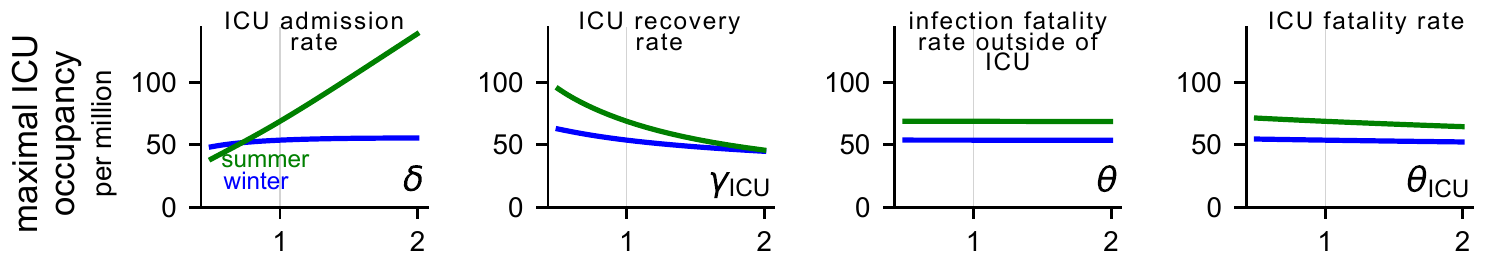}
    \caption{%
        \textbf{Propagation of parameter uncertainties related to age-dependent transition rates.} Parameters are varied independently across their assumed range. Note that the parameters are age-dependent (vector like) and the x-axis indicates a multiplicative factor applied to all values of the vector at the same time, instead of a single averaged parameter value. Therefore, the default parameter value indicated by the grey vertical line is always at 1.}
    \label{fig:sensitivity_2}
\end{figure}

\begin{figure}[!ht]
    \centering
    \includegraphics[width=6in]{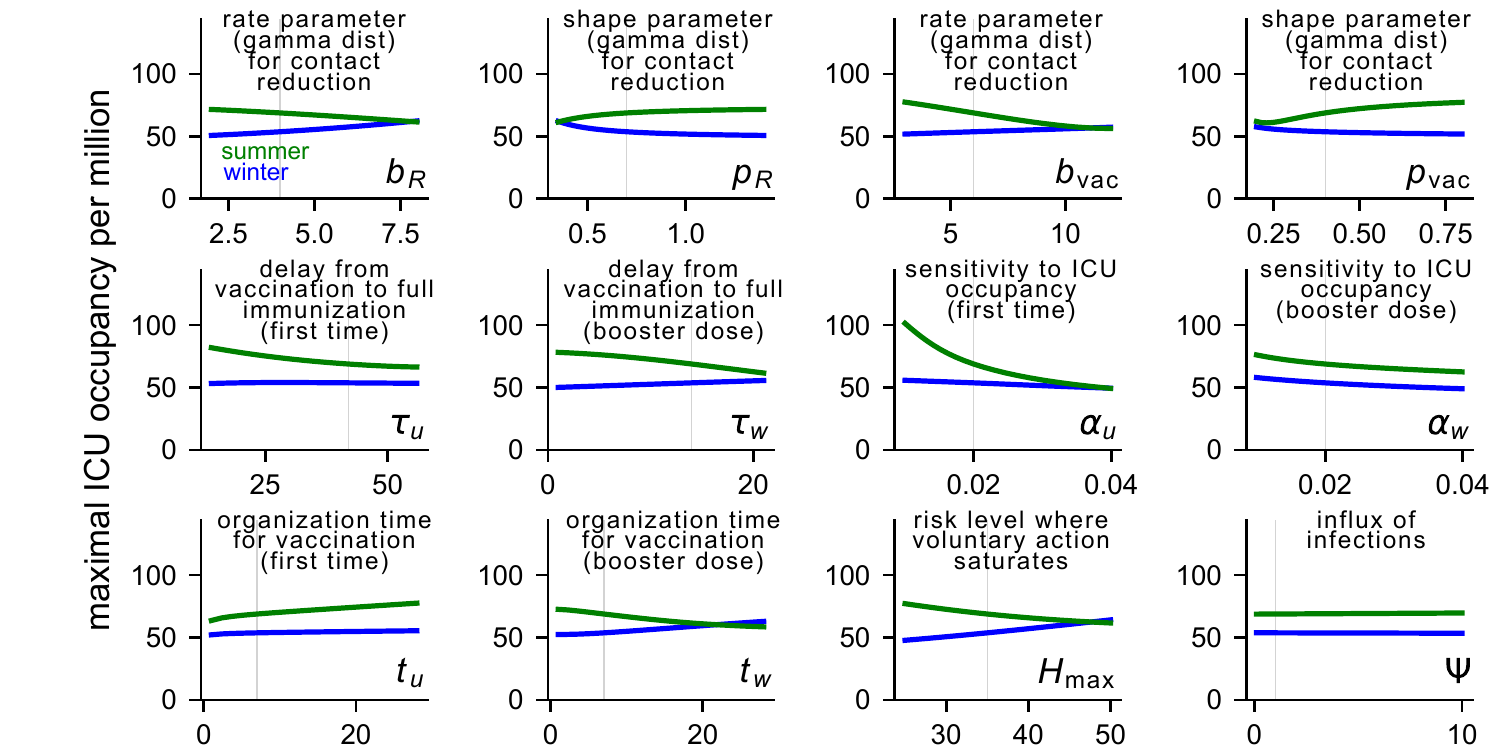}
    \caption{%
        \textbf{Propagation of vaccine uptake- related parameter uncertainties.} Parameters are varied independently across their assumed range. The measured quantity is the maximum ICU occupancy observed in the first wave (blue) and the second wave (green). A vertical line indicates the default value of the parameter. Thus, the points where the green and blue curve intersect the vertical line have the same y-coordinate in all plots.}
    \label{fig:sensitivity_3}
\end{figure}

\begin{figure}[!ht]
    \centering
    \includegraphics[width=6in]{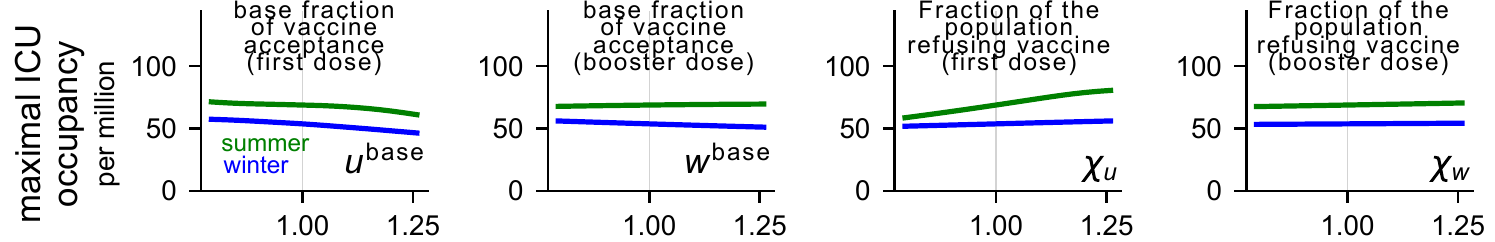}
    \caption{%
        \textbf{Uncertainty propagation of age-dependent parameters related to vaccine uptake.} Parameters are varied independently across their assumed range. Note that the parameters are age-dependent (vector like) and the x-axis indicates a multiplicative factor applied to all values of the vector at the same time, instead of a single averaged parameter value. Therefore, the default parameter value indicated by the grey vertical line is always at 1.}
    \label{fig:sensitivity_4}
\end{figure}

\section{Age-stratified results}

\begin{figure}[!ht]
    \centering
    \includegraphics[width=6in]{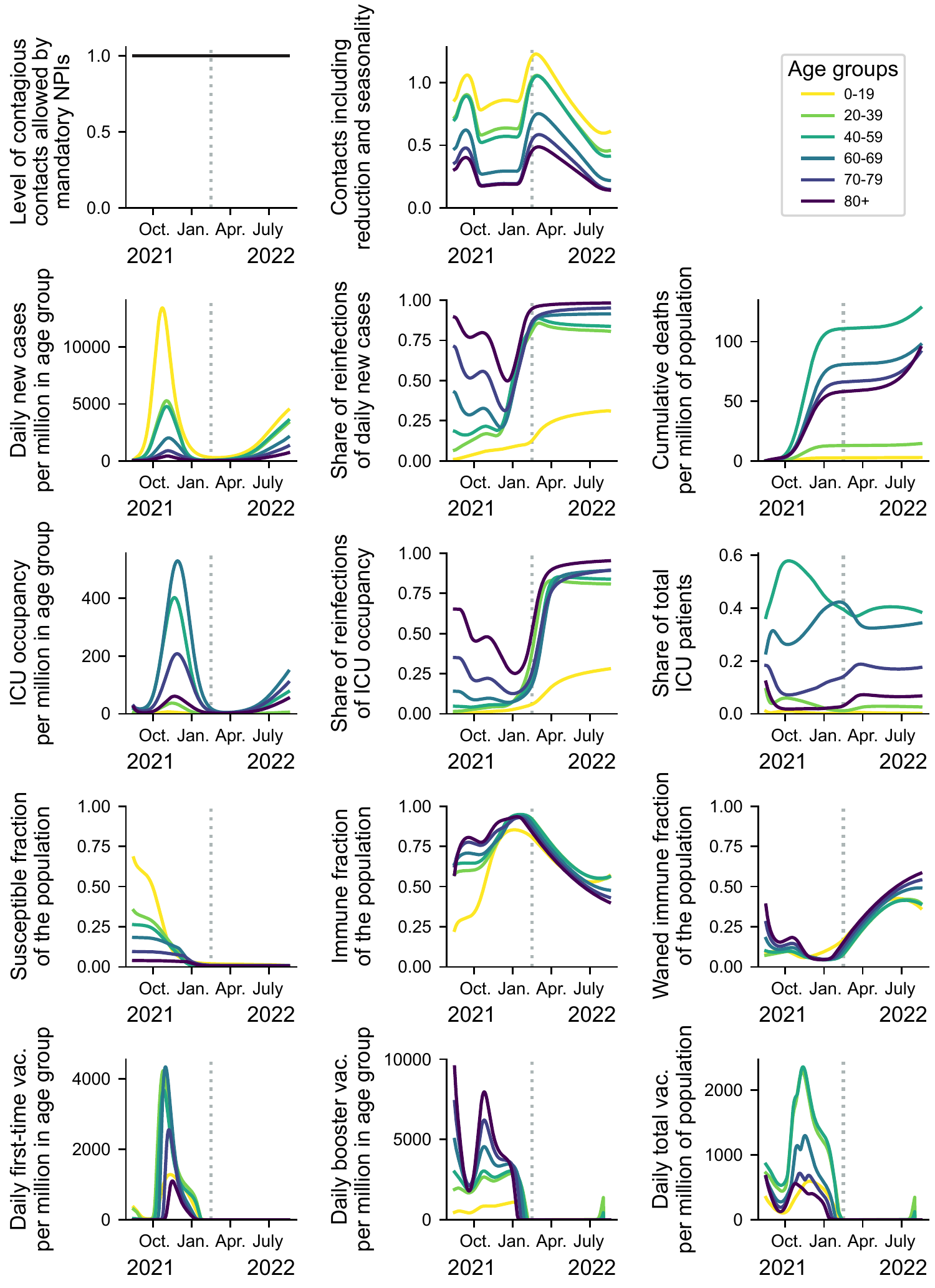}
    \caption{%
        \textbf{Age-stratified results for scenario 1.} }
    \label{fig:AG_1}
\end{figure}

\begin{figure}[!ht]
    \centering
    \includegraphics[width=6in]{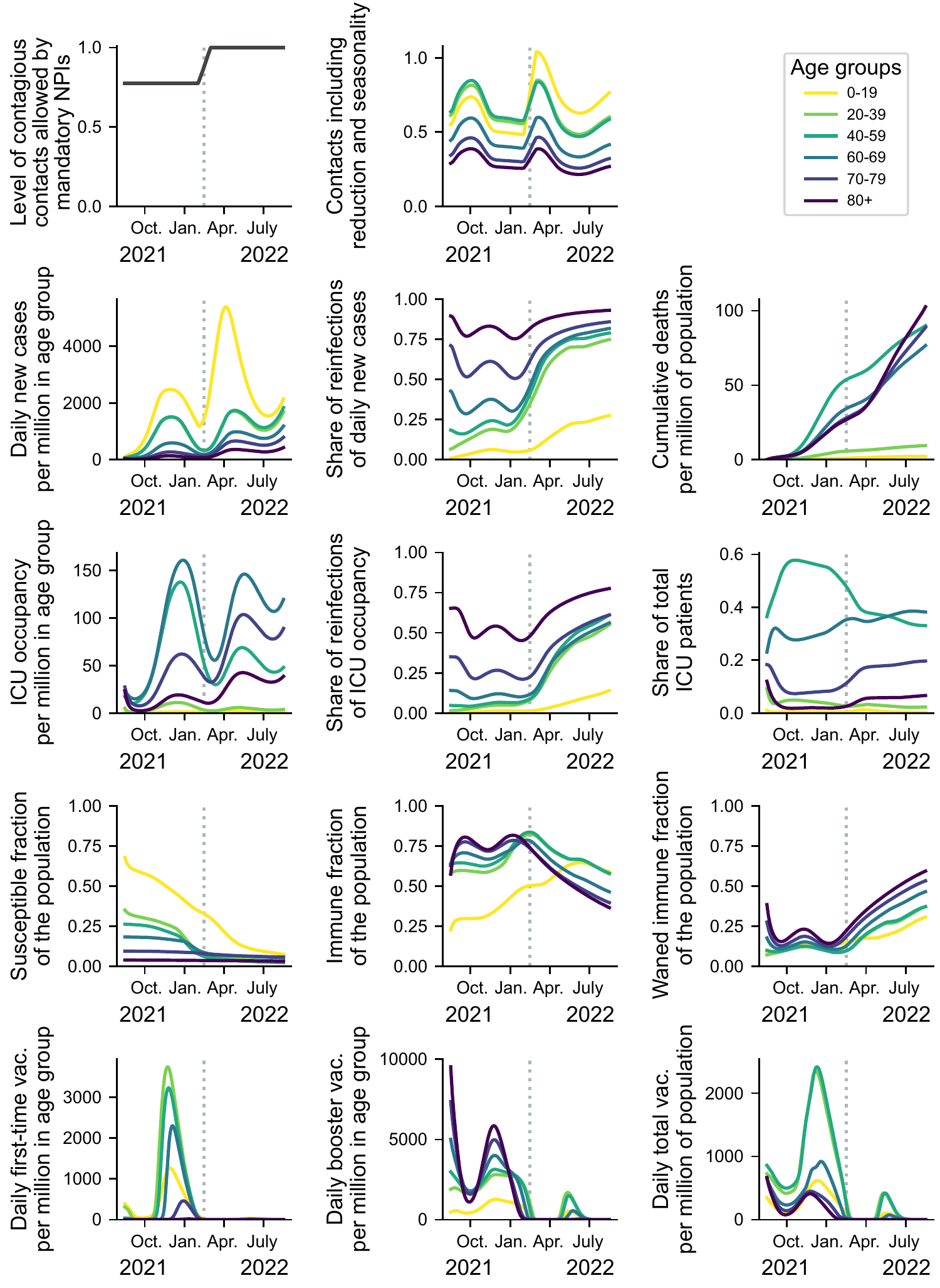}
    \caption{%
        \textbf{Age-stratified results for scenario 2.} }
    \label{fig:AG_2}
\end{figure}

\begin{figure}[!ht]
    \centering
    \includegraphics[width=6in]{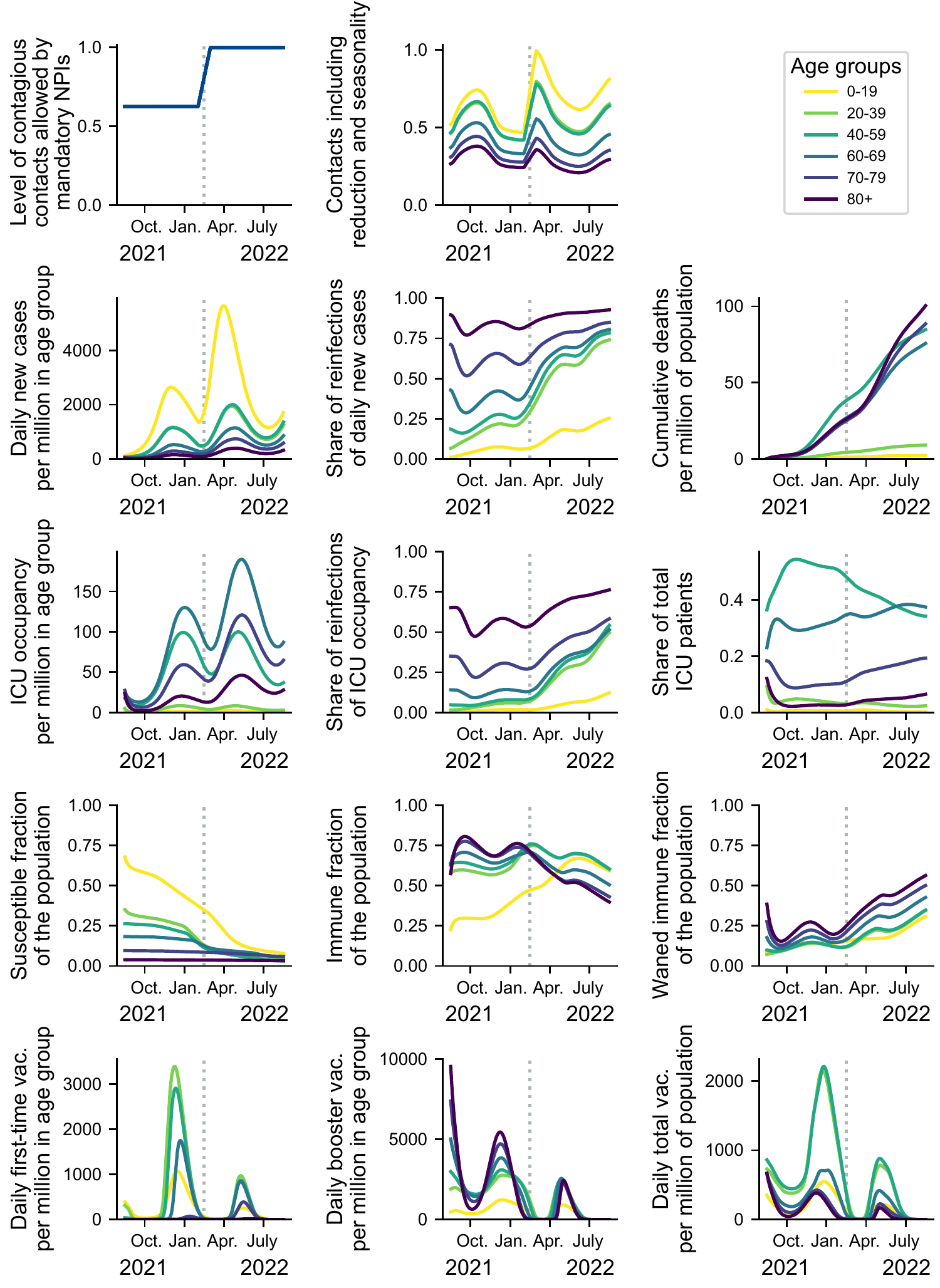}
    \caption{%
        \textbf{Age-stratified results for scenario 3.} }
    \label{fig:AG_3}
\end{figure}

\begin{figure}[!ht]
    \centering
    \includegraphics[width=6in]{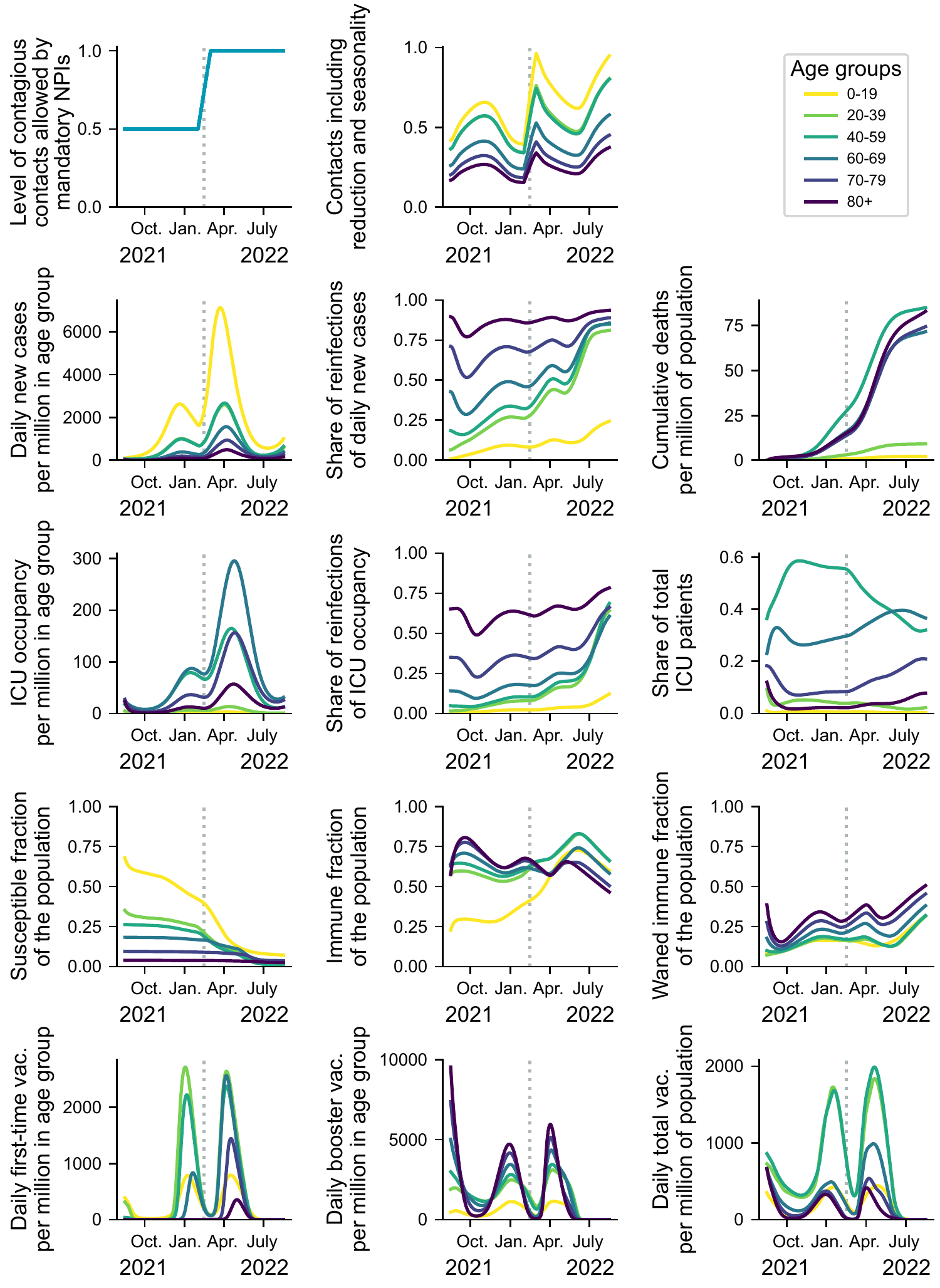}
    \caption{%
        \textbf{Age-stratified results for scenario 4.} }
    \label{fig:AG_4}
\end{figure}

\begin{figure}[!ht]
    \centering
    \includegraphics[width=6in]{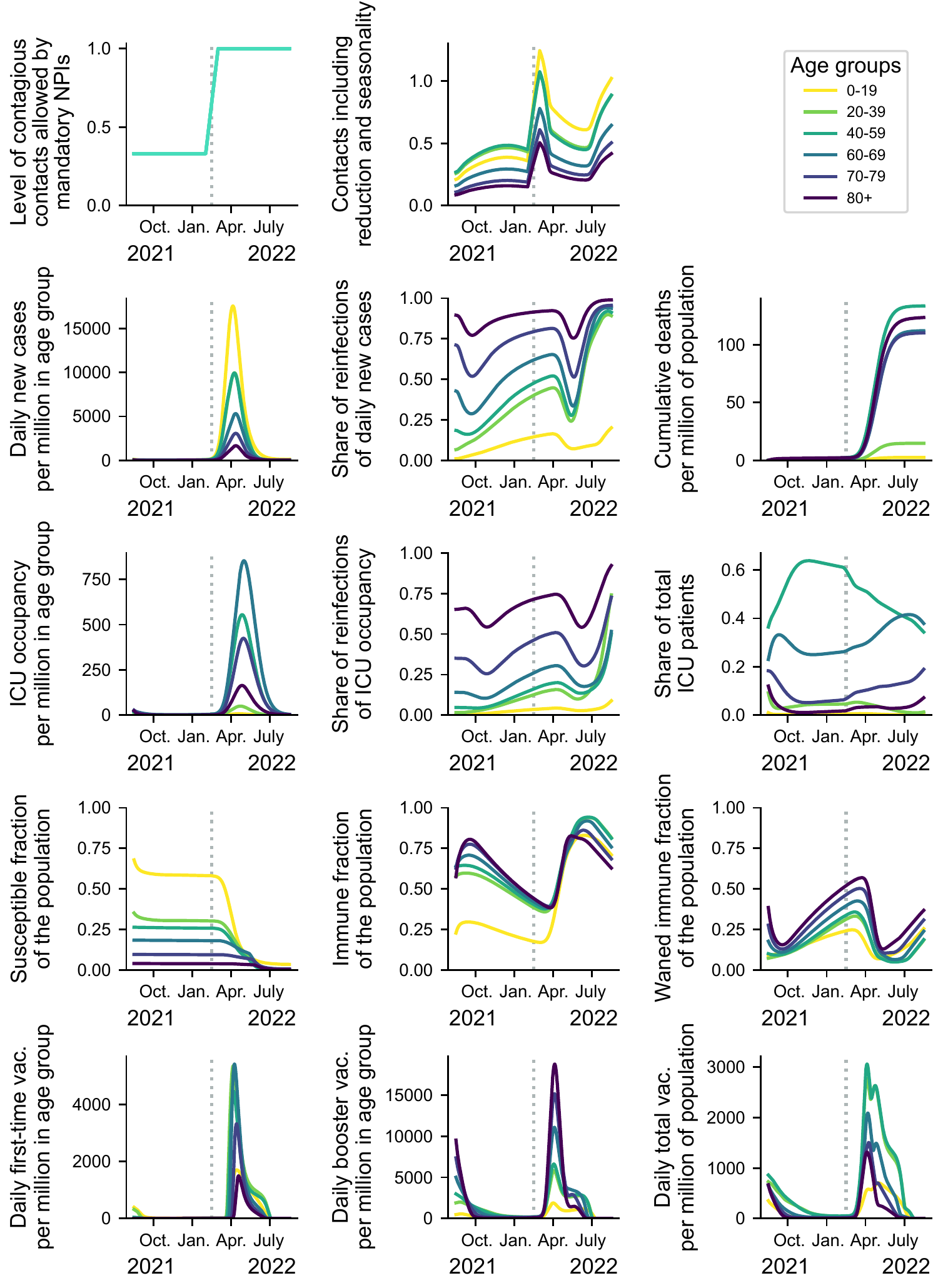}
    \caption{%
        \textbf{Age-stratified results for scenario 5.} }
    \label{fig:AG_5}
\end{figure}

Figures \ref{fig:AG_1}-\ref{fig:AG_5} show the age-stratified results for all scenarios of the main text.

\clearpage

\end{document}